\documentclass[aps,twocolumn,amsmath,amssymb,preprintnumbers]{revtex4}
\usepackage{amsmath} \usepackage{amsfonts} \usepackage{amssymb}
\usepackage{bbm}
\usepackage{epsfig}
\usepackage{graphics}
\usepackage{graphicx}
\newcommand{\be}{\begin{equation}}
\newcommand{\ee}{\end{equation}}
\newcommand{\ba}{\begin{eqnarray}}
\newcommand{\ea}{\end{eqnarray}}

\newcommand{\hmn}{_{\hat\mu\hat\nu}}

\newcommand{\iD}{^{(D)}}
\newcommand{\hmnrs}{_{\hat\mu\hat\nu\hat\rho\hat\sigma}}

\begin{document}

\title[ ]{The cosmological constant and higher dimensional dilatation symmetry}

\author{C. Wetterich}
\affiliation{Institut  f\"ur Theoretische Physik\\
Universit\"at Heidelberg\\
Philosophenweg 16, D-69120 Heidelberg}

\begin{abstract}
We discuss a possible solution to the cosmological constant problem based on the hypothesis of a fixed point for higher dimensional quantum gravity coupled to a scalar. At the fixed point, which is reached only for an infinite value of the scalar field, dilatation symmetry becomes exact. For this limit we concentrate on the absence of a scalar potential since a polynomial potential is not consistent with dilatation symmetry in higher dimensions. We find generic solutions of the higher dimensional field equations for which the effective four-dimensional cosmological constant vanishes, independently of the parameters of the higher dimensional effective action. Under rather general circumstances these are the only quasistatic stable extrema of the effective action which lead to a finite four-dimensional Planck mass. We discuss the associated higher dimensional self-tuning mechanism for the cosmological constant. If cosmological runaway solutions approach the fixed point as time goes to infinity, the effective dark energy vanishes asymptotically.  In the present cosmological epoch the fixed point is not yet reached completely, resulting in a tiny amount of dark energy, comparable to dark matter. We discuss explicitly higher dimensional geometries which realize such asymptotic solutions for time going to infinity.
\end{abstract}

\maketitle

\section{Introduction}
\label{intro}
Anomalous dilatation symmetry may be a key ingredient for a dynamical solution of the cosmological constant problem \cite{CWQ}, \cite{PSW}. In the asymptotic limit of time $t$ going to infinity, a cosmological runaway solution of the field equations can approach a fixed point. At the fixed point all memory of explicit mass or length scales is lost and the quantum effective action becomes dilatation symmetric. In other words, the dilatation anomaly vanishes when it is evaluated for the field configurations corresponding to the fixed point \cite{CWCC}. In consequence, a scalar field becomes massless in the asymptotic limit for $t\to\infty$, corresponding to the Goldstone boson of spontaneously broken dilatation symmetry. 

For the approach to the fixed point at finite $t$ the anomaly is not yet zero, and correspondingly the scalar ``pseudo-Goldstone boson'' still has a small mass, that vanishes only asymptotically. These ideas are realized in practice in quintessence cosmologies, where the ``cosmon''-field plays the role of the pseudo-Goldstone boson of spontaneously broken anomalous dilatation symmetry \cite{CWQ}, \cite{CWCC}. The cosmon mass is varying with time and of the order of the Hubble parameter \cite{CWAA}.

Before discussing cosmological solutions, it is crucial to understand the asymptotic solution towards which the cosmological runaway solution converges asymptotically. In our scenario, this asymptotic solution should correspond to flat Minkowski space and therefore to an ``asymptotically vanishing cosmological constant''. However, a vanishing cosmological constant is not enforced by dilatation symmetry - asymptotic dilatation symmetry could, in principle, also be realized with a non-zero effective cosmological constant.

In two recent papers \cite{CWCC}, \cite{CWNL} we have argued that a higher dimensional setting sheds new light on the question why the effective four-dimensional cosmological constant $\Lambda$ vanishes asymptotically. Actually, such an asymptotic vanishing happens for a large generic class of cosmological runaway solutions without any tuning of parameters. The remaining problem concerns the strict limits on the possible variation of fundamental couplings, which are not obeyed for some classes of such solutions. In ref. \cite{CWCC} we have discussed several scenarios how to reconcile asymptotically static couplings (with an interesting possible exception in the neutrino sector \cite{CWN}) and an asymptotically vanishing cosmological constant.

In the present paper we investigate higher dimensional models for which the quantum effective action exhibits an exact dilatation symmetry for a suitable fixed point. This fixed point should become relevant for the asymptotic solution. A dilatation symmetric effective action is also the starting point for approaches where dilatation symmetry is realized as an exact quantum symmetry \cite{SD}. Thus the basic object of our investigation is the quantum effective action $\Gamma$ where all quantum fluctuations are already included. The field equations derived from $\Gamma$ are exact without any further quantum corrections. Our key finding is that in the presence of higher dimensional dilatation symmetry these field equations have generic solutions for which the four-dimensional cosmological constant vanishes, $\Lambda=0$. They correspond to stable extrema of $\Gamma$.

We do not postulate here that the effective action of a fundamental theory is dilatation symmetric. In general, it is not, and dilatation anomalies are present. For example, a dilatation anomaly arises if running couplings induce an explicite scale, or if a higher dimensional cosmological constant is present. We only make the hypothesis that $\Gamma$ has a fixed point for certain asymptotic field values to be specified below. Only in this asymptotic limit the dilatation anomaly vanishes - only the ``fixed point effective action'' is dilatation symmetric.

For a cosmological runaway solution the values of fields are not static but continue to change for all times. We investigate solutions where the fields move towards the fixed point region for $t\to\infty$. For $t\to\infty$ the field equations derived from the fixed point effective action become accurate. As the runaway solution approaches a fixed point the anomalous parts of the effective action vanish. Since such a fixed point is generally reached only asymptotically for $t\to\infty$, it is only in this limit that the field equations exhibit an exact dilatation symmetry. 

Our discussion of a dilatation symmetric effective action will therefore be insufficient for the cosmological solutions. The solutions of the field equations derived from the dilatation symmetric effective action account only for the possible asymptotic states for $t\to\infty$. Nevertheless, it is a necessary condition in our scenario that the dilatation symmetric effective action has an extremum which describes an acceptable asymptotic state. This solution should have flat four-dimensional space. Furthermore, a simple solution to the problem of time-varying couplings would be a static ratio between the characteristic length scale of internal geometry and the effective four-dimensional Planck length. The latter should lead after dimensional reduction to finite non-zero values of the gauge couplings and other dimensionless couplings and mass ratios in the resulting model of particle physics. Inversely, if a satisfactory asymptotic solution exists, the chances that a runaway solution will approach it for $t\to\infty$ are quite high.

Dilatation symmetry is easily realized in models that contain besides the metric also a scalar field $\xi$. A striking difference between dilatation symmetry in higher dimensions, $d>6$, as compared to four dimensions is the absence of a polynomial potential for a scalar field with a canonical kinetic term \cite{CWCC}. Indeed, for $d=4~mod~4$ (or $d=2~mod~4$ with a symmetry $\xi\to-\xi$) the most general effective action , which is polynominal in $\xi$ and consistent with general coordinate transformations (diffeomorphism-symmetry) and global dilatation symmetry, reads
\begin{equation}\label{AA1a}
\Gamma=\int_{\hat x}\hat g^{1/2}\Big\{-\frac12\xi^2\hat R
+\frac\zeta2\partial^{\hat\mu}\xi\partial_{\hat\mu}\xi
+F(\hat R_{\hat\mu\hat\nu\hat\rho\hat\sigma})\Big\}.
\end{equation}
The normalization of $\xi$ is chosen such that it has a canonical kinetic term, up to the free dimensionless parameter $\zeta$.   While for $d=4$ a term $\lambda\xi^4$ is dilatation invariant (and for $d=6$ a term $\gamma\xi^3$), no polynomial $\xi^n$ with integer $n$ is dilatation invariant for $d>6$. 

We may first investigate the case that $F$ contains only polynomials of the curvature tensor $\hat R_{\hat\mu\hat\nu\hat\rho\hat\sigma}$ or derivatives thereof. A typical example for $F$ is
\begin{equation}\label{AA2a}
F=\tau\hat R^{\frac d2},
\end{equation}
and we will discuss more general polynomial forms of $F$ in the next section (cf. ref. \cite{CWCC}, \cite{CWNL}.) For odd dimensions no invariant polynomial in the curvature tensor exists and therefore $F=0$. (We disregard invariants involving the $\epsilon$-tensor.) We will often concentrate on the simplest form of such a fixed point effective action by taking $F=0$ in eq. \eqref{AA1a}. Nevertheless, we also discuss more general forms. The polynomial form of $F$ is actually not essential for our findings and we will extend our discussion to the most general dilatation symmetric term $F$ that only involves the metric. In contrast, the absence of a dilatation symmetric (non-polynomial) potential $V(\xi)$ at the fixed point remains important for certain aspects of our discussion. We only briefly comment in sect. \ref{quasi} on possible modifications if a non-polynominal potential for $\xi$ would be present at the fixed point. It seems likely that this will not change the existence of extrema of $\Gamma$ with $\Lambda=0$.

The key argument for the generic existence of solutions with $\Lambda=0$ is rather simple and directly related to higher dimensional dilatation symmetry. Let us consider the most general form of a dilatation symmetric quantum effective action
\begin{equation}\label{2A}
\Gamma=\int_{\hat x}\hat g^{1/2}{\cal L}.
\end{equation}
Dilatation transformations correspond to a rescaling of the metric by a constant factor $\alpha^2$, and an associated rescaling of $\xi$,
\begin{eqnarray}\label{2B}
\hat g_{\hat\mu\hat\nu} \to \alpha^2\hat g_{\hat\mu\hat\nu}~,~\hat g^{1/2}\to\alpha^d\hat g^{1/2},\nonumber\\
\xi\to\alpha^{-\frac{d-2}{2}}\xi~,~{\cal L}\to\alpha^{-d}{\cal L}.
\end{eqnarray}
A dilatation symmetric effective action remains invariant under these rescalings. For general field values we may define 
$\Gamma_\kappa[\hat g_{\hat\mu\hat\nu},\xi]=\Gamma[\kappa^{-2}\hat g_{\hat\mu\hat\nu},
\kappa^{\frac{d-2}{2}}\xi]$ and consider the asymptotic limit $\kappa\to\infty$. Our fixed  point hypothesis states then that $\Gamma_\kappa$ becomes dilatation symmetric for $\kappa\to\infty$. 

The special role of dilatation symmetry for the problem of the cosmological constant is visible already for the most general form of a dilatation symmetric effective action. We are interested in configurations with a block diagonal metric
\begin{equation}\label{2c}
\hat g_{\hat\mu\hat\nu} (x,y)=
\left(\begin{array}{ccc}
\sigma(y)g^{(4)}_{\mu\nu} (x)
&,&0\\
0&,&g^{(D)}_{\alpha\beta}(y)
\end{array}\right).
\end{equation}
Here $x^\mu$ denotes the four-dimensional coordinates and $y^\alpha$ are coordinates of $D$-dimensional internal space, with corresponding metrics $g^{(4)}_{\mu\nu}$ and $g^{(D)}_{\alpha\beta},d=D+4$. The function $\sigma(y)$ accounts for a possible warping \cite{RSW,7A,RDW,RS}. The configuration for the metric is supplemented by a configuration for the scalar field $\xi(y)$. With 
$\hat g^{1/2}=(g^{(4)})^{1/2}\sigma^2(g^{(D)})^{1/2}$ we define
\begin{equation}\label{2D}
W(x)=\int_y(g^{(D)}{(y)})^{1/2}\sigma^2(y){\cal L}(x,y),
\end{equation}
and write
\begin{equation}\label{2E}
\Gamma=\int_x(g^{(4)})^{1/2} W.
\end{equation}
We observe the scaling under dilatations $(g^{(4)}_{\mu\nu}\to\alpha^2 g^{(4)}_{\mu\nu}~,~g^{(D)}_{\alpha\beta}\to\alpha^2 g^{(D)}_{\alpha\beta})$
\begin{equation}\label{2F}
W\to\alpha^{-4}W.
\end{equation}

The possible extrema of $\Gamma$ can be classified into two categories, according to the existence of an extremum of $W(x)$ or not. Consider first the case where an extremum of $W(x)$ exists for an appropriate field configuration $\hat g_{\hat\mu\hat\nu},\xi$. This means that a (infinitesimally) close neighboring field configuration does not change the function $W(x)$. We may denote the value of $W(x)$ at its extremum by $W_0(x)$. The scaling property \eqref{2F} immediately implies $W_0(x)=0$. Indeed, using in eq. \eqref{2B} $\alpha=1+\epsilon$, with infinitesimal $\epsilon$, defines a neighboring configuration. The combination of scaling and extremum condition, $\partial_\epsilon(1+\epsilon)^{-4}W_0=0$, can be obeyed only for $W_0=0$. With eq. \eqref{2E} and $W_0=0$ an extremum of $W$ is also an extremum of the effective action $\Gamma$. Thus, whenever an extremum of $W$ exists, this defines an extremum of $\Gamma$ obeying the field equations, with a vanishing value of $\Gamma$ at the extremum, $\Gamma_0=0$. 

For all solutions which admit dimensional reduction to an effective four dimensional ``local'' theory of gravity we will show that $\Gamma_0=0$ implies that the effective four-dimensional cosmological constant $\Lambda$ vanishes. The existence of extrema of $W$ is a rather generic feature and we find that within our setting stable extrema in this ``flat phase'' always exist. Only the question remains if this flat phase includes solutions with interesting particle physics. There may exist additional extrema of $\Gamma$ which are not extrema of $W$ and for which $\Lambda$ does not vanish. Four dimensional theories with a non-vanishing cosmological constant are possible only for this possible second category of extrema of $\Gamma$ which are not extrema of $W(x)$. In the absence of a non-polynominal scalar potential we find that all such possible solutions in the non-flat phase are unstable if $\xi \neq 0$. This singles out a vanishing four-dimensional cosmological constant.

The  existence of extrema of $W$ is a qualitative feature  that does not depend on the precise values of the parameters characterizing ${\cal L}$. As an example we may consider the polynomial effective action \eqref{AA1a}. It is obvious that $\hat R_{\hat\mu\hat\nu\hat\rho\hat\sigma}=0~,~\xi=\xi_0=$const. corresponds to such an extremum, with ${\cal L}_0=0$. The existence of this extremum extends to a very large class of non-polynomial ${\cal L}$ as well. It is sufficient that the $\xi$-dependent part of ${\cal L}$ vanishes for flat space and constant $\xi$, and that the remaining $\xi$ independent part $F$ vanishes for $\hat R_{\hat\mu\hat\nu\hat\rho\hat\sigma}=0$. 

Furthermore, if
\begin{equation}\label{2G}
W_F(x)=\int_y(g^{(D)})^{1/2}\sigma^2 F
\end{equation}
admits an extremum with respect to variations of the metric, this must occur for $W_{F,0}(x)=0$ by the same scaling arguments as above. (No polynomial form of $F$ needs to be assumed here.) A dilatation symmetric theory involving only gravity (without the scalar field $\xi$) will imply a vanishing cosmological  constant if an extremum of $W_F$ exists which leads to acceptable four dimensional gravity (with nonzero and finite effective Planck mass). 

Let us next extend the setting by adding a scalar part ${\cal L}_\xi$ which is quadratic in $\xi$, without being necessarily polynomial in the curvature tensor. For a given metric $\hat g_{\hat\mu\hat\nu}$ we can find a partial extremum of $\Gamma_\xi=\int_{\hat x}\hat g^{1/2}{\cal L}_\xi=\int_x(g^{(4)})^{1/2} W_\xi$ by solving the higher dimensional field equation for $\xi$, consistent with an extremum condition in case of singular geometries which will be derived later. This results in $W_\xi=0$, as expected for possible extrema of a purely quadratic polynomial. A necessary condition for an extremum of $W=W_\xi+W_F$ remains therefore $W_F=0$. However, only the sum $W_\xi+W_F$ has to be an extremum with respect to variations of the metric, and not $W_\xi$ and $W_F$ separately. We discuss an example of this type of extrema in the appendix. We observe that for a quadratic ${\cal L}_\xi$ we could ``integrate out'' the scalar field in favor of non-local gravitational interactions. This demonstrates incidentally that our setting is not restricted to a local effective gravitational action. 

From these simple observations we conclude that in the presence of dilatation symmetry the vanishing of the cosmological constant $\Lambda$ is very robust with respect to variations of the precise form of $\Gamma$. The effective action may be characterized by many parameters, as for example the coefficients of different terms appearing in a polynomial approximation of $F$. A change of the values of these parameters will typically not change the value $\Lambda=0$, as long as an extremum of $W$ continues to exist and is compatible with effective four dimensional gravity. This feature is a particular consequence of dilatation symmetry - it no longer holds in presence of dilatation anomalies. In view of the robustness of $\Lambda=0$ we find many analogies to phases in many body theories. For the ``flat phase'' an extremum of $W$ exists and $\Lambda=0$. Extrema of $\Gamma$ in the ``non-flat phase'' are not extrema of $W$, and we will typically find $\Lambda>0$. 

For effective four dimensional theories in the flat phase we will show that $W$ can be associated with the effective potential $V$ for four-dimensional scalar fields. The minimum of $V$ occurs then necessarily for $V_0=0$. Furthermore, $V$ has flat directions. One such flat direction corresponds to the dilatations \eqref{2B} and the corresponding massless scalar field is the dilaton. There may be more flat directions since the extrema of $W(x)$ may occur for a whole class of different configurations $\hat g_{\hat\mu\hat\nu},\xi$. In this case we expect additional massless fields for the dilatation symmetric asymptotic cosmological solution for $t\to\infty$. For finite $t$ the presence of dilatation anomalies will induce mass terms for these scalars, which vanish only for $t\to\infty$. It is conceivable that such additional light scalar fields (beyond the cosmon) are interesting candidates for dark matter.

Our investigation focuses on two issues.
\begin{itemize}
\item [(i)] The existence of extrema of $W$ and a first discussion of characteristic properties of solutions in the flat phase.
\item [(ii)] The dimensional reduction to effective four dimensional gravity  and the establishment that $\Lambda=0$ in the flat phase.
\end{itemize}
A demonstration that an effective action which admits a flat phase does generically not allow other extrema with arbitrarily small $|\Lambda|$ is given in ref. \cite{CWNL}. In particular, there are no continuous families of extrema where $\Lambda$ appears as a continuous parameter. This issue is important for warped geometries with singularities where the existence of families of solutions of the higher dimensional field equations with continuous $\Lambda$ is known \cite{RSW}, \cite{7A}, \cite{RDW}.  In this case the extremum conditions for $\Gamma$ go beyond the higher dimensional field equations \cite{CWCON}. They precisely select the solutions with $\Lambda=0$ out of the continuous family of solutions. The particular properties of warped spaces within our general dilatation symmetry setting will be discussed in an accompanying paper \cite{P2}.

In the course of our discussion we will explicitly address the issues of quantum corrections, ``tuning of the cosmological constant to zero'', and ``naturalness'' of the solutions with $\Lambda=0$. Some of the general aspects are already discussed in \cite{CWCC} and not repeated here, such that we concentrate here on our specific setting. We discuss the special role of higher dimensions for the ``self-tuning'' of the cosmological constant to zero. This self-tuning is particular to the case of dilatation symmetry and closely connected to the robustness of the existence of extrema of $W$ under a change of parameters in the effective action. We find rather satisfactory answers to the naturalness problem. Asymptotic dilatation symmetry in higher dimensional theories may indeed provide the key for a solution of the cosmological constant problem. 

Our paper is organized as follows. In sect. \ref{dilatation} we present a first discussion of a polynomial effective action \eqref{AA1a}. We establish explicitely the existence of extrema of $W$ belonging to the flat phase for a large class of effective actions, and demonstrate the vanishing of the cosmological constant if effective four dimensional gravity exists. We extend this discussion to the most general dilatation symmetric effective action and show that extrema in the flat phase with $\Lambda=0$  exist whenever an appropriate $d-4$ dimensional functional $\bar W$ admits an extremum. Section \ref{quasi} addresses the existence of the two ``phases'' for solutions in an effective four dimensional framework and shows why $\Lambda=0$ is singled out in the flat phase. We concentrate on ``quasistatic solutions'' for which the four dimensional fields are static and homogeneous, while the metric describes a geometry with maximally four dimensional symmetry, with positive, negative or vanishing $\Lambda$. It also demonstrates that the non-flat phase with $\Lambda>0$ can only be realized if certain conditions in parameter space are met. The non-flat phase appears to be less generic than the flat phase. We find that all possible extrema in the non-flat phase for $\xi \neq 0$ are unstable.

In sect. \ref{extended} we discuss an interesting ``extended scaling symmetry'' which becomes realized for the simplest fixed point with $F=0$, or if the contribution of the term $\sim F$ can be neglected for the extremum condition. Extended scaling symmetry admits quasistatic solutions only in the flat phase. Sect. \ref{adjustment} deals with the issue of an ``adjustment'' or tuning'' of the cosmological constant. The robustness of $\Lambda=0$ with respect to changes of parameters in the dilatation symmetric effective action is associated to a mechanism of ``self-adjustment'' or ``self-tuning''. In this respect we highlight the differences between a four dimensional theory with a finite number of scalar fields and a higher dimensional setting with infinitely many effective four dimensional scalar fields. This is an important ingredient for the understanding of the robustness of the flat phase, which would be hard to implement with a finite number of degrees of freedom. 

Sect. \ref{higherorder} investigates the dilatation symmetric effective action \eqref{AA1a} with a polynomial form of $F$. We display the field equations and discuss simple solutions with a vanishing four-dimensional cosmological constant $\Lambda$. An appendix is devoted to particular solutions with non-Ricci-flat internal space, which nevertheless results in $\Lambda=0$. These solutions demonstrate explictly that non-abelian isometries of internal space are possible for extrema in the flat phase.

The final part presents a simple estimate of the dilatation anomaly in sect. \ref{dilatationanomaly}. In sect. \ref{cosmologicalrunaway} we show that this leads to cosmological runaway solutions for which $\Lambda$ vanishes asymptotically, while for finite $t$ a homogeneous dark energy component accounts for quintessence. We present our conclusions in sect. \ref{conclusions}.

\section{Dilatation symmetry and vanishing four-dimensional cosmological constant}
\label{dilatation}

In this section we show a few striking general properties of solutions to the higher dimensional field equations which are derived from a dilatation symmetric polynomial effective action \eqref{AA1a}. (i) There exist always solutions for which the four-dimensional effective cosmological constant $\Lambda$ vanishes. (ii) Particular solutions of this type, with constant $\xi=\xi_0$ and $\hat R_{\hat\mu\hat\nu\hat\rho\hat\sigma}=0$, exist for arbitrary polynomial $F$. The existence of these solutions with $\Lambda=0$ extends to non-polynomial $F$, provided $F$ is analytic around flat space, $\hat g_{\hat\mu\hat\nu}=\eta_{\hat\mu\hat\nu}$ or, even weaker, provided an expansion of $\int\hat g^{1/2}F$ linear in $\hat h_{\hat\mu\hat\nu}=\hat g_{\hat\mu\hat\nu}-\eta_{\hat\mu\hat\nu}$ exists and the coefficient of the linear term vanishes. (Typically, $F$ is at least quadratic in $\hat h_{\mu\nu}$.) (iii) Further solutions with $\Lambda=0$ and $\hat R_{\hat\mu\hat\nu\hat\rho\hat\sigma}\neq 0$ are possible. For example, if $F(\hat R=0)=0$, all solutions with $\xi=\xi_0$ must have $\Lambda=0$.

\medskip\noindent
{\bf 1. Absence of dilaton potential and vanishing

\hspace{0.06cm}  cosmological constant}

The particular features arising from the absence of a potential for the scalar field $\xi$ and their impact on the cosmological constant problem can be realized by the following chain of arguments. First, for any solution with a constant non-zero value of $\xi$ the curvature scalar $\hat R$ must vanish. This follows from the scalar field equation
\begin{equation}\label{AA3a}
\xi\hat R=-\zeta\hat D^2\xi.
\end{equation}
In contrast, for $d=4$ eq. \eqref{AA3a} would contain an additional term $4\lambda\xi^3$, such that for $\xi=const.$ one would infer a nonzero $\hat R$ for $\lambda\neq 0$, i.e. $\hat R=4\lambda\xi^2$. Second, if $F$ vanishes for $\hat R=0$, as for the example \eqref{AA2a}, one infers that for solutions with constant $\xi$ the effective action must vanish, $\Gamma=0$. Third, let us consider solutions with maximal four-dimensional symmetry, i.e. Poincar\'e-symmetry for flat space or the corresponding symmetries of de Sitter or anti-de Sitter space. (The corresponding geometries have a zero, positive or negative cosmological constant, respectively.) For this class of solutions a vanishing effective action, $\Gamma=0$, implies flat space or a vanishing four-dimensional cosmological constant. The last step follows for any solution which admits dimensional reduction to an effective four-dimensional theory. This should generically be the case if the geometry of the ``internal dimensions'' exhibits a characteristic length scale $l$ (``compactification scale''), such that for present observations with ``large wavelengths'' as compared to $l$ internal space remains unobservable. 

\medskip\noindent
{\bf 2. Vanishing extremum value of $\Gamma$ implies $\Lambda=0$}

Let us develop the argument why extrema of the action with $\Gamma_0=0$ imply $\Lambda=0$ in some detail. Within higher dimensional theories with a small characteristic length scale an effective four dimensional action $\Gamma^{(4)}$ can be obtained by expanding the higher-dimensional fields $\xi$ and $\hat g_{\hat\mu\hat\nu}$ in a complete system of functions in internal space, and subsequently integrating over internal space. The coefficients of the expansion correspond to infinitely many four dimensional fields, often comprising massless fields and a tower of massive Kaluza-Klein modes. If the higher dimensional action vanishes for a given field configuration $\hat g\hmn,\xi$, the four dimensional effective action has to vanish as well, $\Gamma^{(4)}=0$. Here $\Gamma^{(4)}$ is evaluated for values of the four-dimensional fields that correspond to the given higher dimensional configuration. 

Furthermore, if the higher dimensional fields are unconstrained, any extremum of the higher dimensional action must be an extremum of $\Gamma^{(4)}$ with respect to the variation of each four-dimensional field. In particular, $\Gamma^{(4)}$ must be an extremum with respect to the variation of the four-dimensional metric $g^{(4)}_{\mu\nu}$. (We indicate higher dimensional quantities and indices by a hat, $\hat\mu,\hat\nu,\dots$, and four dimensional indices without hat, $\mu,\nu,\dots$) For any extremum of $\Gamma$ we can therefore set all four dimensional fields to the values corresponding to this extremum, except for $g^{(4)}_{\mu\nu}$. Evaluating the four-dimensional action for this situation reduces $\Gamma^{(4)}[g^{(4)}_{\mu\nu}]$ to a functional depending only on $g^{(4)}_{\mu\nu}$. If $\Gamma$ is an extremum, $\Gamma^{(4)}$ must be an extremum with respect to the variations of $g^{(4)}_{\mu\nu}$. (See ref. \cite{CWCC} for a more extended discussion.) In other words, $g^{(4)}_{\mu\nu}$ has to obey the four-dimensional field equations.

We now consider the most general form of $\Gamma^{(4)}[g^{(4)}_{\mu\nu}]$ in the case where a derivative expansion is valid on cosmological scales and where possible four-dimensional scalar fields take constant values
\begin{equation}\label{AA4a}
\Gamma^{(4)}=\int_x(g^{(4)})^{1/2}
\left\{V-\frac{\chi^2}{2}R^{(4)}+\dots\right\}.
\end{equation}
This should be a valid approximation whenever four-dimensional gravity is effectively local, which we consider to be a generic case. The field equations corresponding to the extremum condition for $\Gamma^{(4)}$ are Einstein's equations with a cosmological constant
\begin{equation}\label{AA5a}
\chi^2\left(R^{(4)}_{\mu\nu}-\frac12 R^{(4)}g^{(4)}_{\mu\nu}\right)=-V g^{(4)}_{\mu\nu}.
\end{equation}
Contraction of this equation, $R^{(4)}=4V/\chi^2=4\Lambda$, and insertion into eq. \eqref{AA4a}, yields the value of $\Gamma^{(4)}$, and therefore also the higher dimensional action $\Gamma$, at the extremum
\begin{equation}\label{AA6a}
\Gamma=\Gamma^{(4)}=-V\int_x(g^{(4)})^{1/2}.
\end{equation}
A vanishing value of $\Gamma$ at the extremum implies a vanishing four-dimensional cosmological constant $V=\chi^2\Lambda=0$, and vice versa. This closes the argument that $\Gamma_0=0$ implies $\Lambda=0$.  In a higher dimensional context, a solution to the cosmological constant problem amounts to finding a quantum effective action which vanishes precisely at its extremum, without fine tuning of parameters. This should hold at least asymptotically for $t\to\infty$.

\medskip\noindent
{\bf 3. Simple higher-dimensional solutions with $\Lambda=0$}

We next discuss a few simple cases where extrema of $\Gamma$ imply $\Lambda=0$, independently of the precise choice of couplings parameterizing $\Gamma$. 

For an effective action with dilatation symmetry we have seen that an asymptotic solution with $\partial_{\hat\rho}\xi=0$ implies $\hat R=0$. In turn, this implies $\Gamma=0$ whenever $F$ contains only terms that involve at least one power of $\hat R$ or its covariant derivatives. A dilatation symmetric polynomial of $\hat R_{\hat\mu\hat\nu\hat\sigma\hat\lambda}$ or its covariant derivatives can always be written in the form $F=F_R+F_H$, where $F_R$ vanishes for $\hat R=0$, and $F_H$ involves only the traceless tensors 
\begin{equation}\label{AA7a}
\hat H_{\hat\mu\hat\nu}=\hat R_{\hat\mu\hat\nu}-\frac1d\hat R\hat g_{\hat\mu\hat\nu}~,~\hat C_{\hat\mu\hat\nu\hat\rho\hat\sigma}
=\hat R_{[\hat\mu\hat\nu\hat\rho\hat\sigma]}
\end{equation}
or their covariant derivatives. (The symbol $[\hat\mu\hat\nu\hat\rho\hat\sigma]$ stands for total antisymmetrization in all indices.) A sufficient condition for a solution of the higher dimensional field equations with constant $\xi$ and a vanishing effective four-dimensional cosmological constant $\Lambda$ is therefore that $F_H$ vanishes for this solution, $F_{H_0}=0$. In particular, we may discuss the class of possible fixed point actions with $F_H=0$. This encloses a large variety of actions, parameterized by many dimensionless couplings. For example, such couplings can be associated with the coefficients of dilatation invariant polynomials with at least one factor $\hat R$. For $F_H=0$, all solutions of the higher dimensional field equations with constant $\xi$ will lead to $\Lambda = 0$.

Furthermore, if $F_R$ is at least quadratic in $\hat R$ and $F_H$ involves at least two powers of $\hat H_{\hat\mu\hat\nu}$ or its covariant derivatives, any configuration with $\partial_{\hat\rho}\xi=0~,~\hat R_{\hat\mu\hat\nu}=0$ will solve the higher dimensional field equations, with $\Gamma_0=0$ for the solution. Independently of the values of the various dimensionless couplings appearing in $F_R$ or $F_H$, all extrema of $\Gamma$ will lead to a vanishing four-dimensional cosmological constant. Dilatation symmetric effective actions of this type always admit solutions in the flat phase, and therefore with $\Lambda=0$.

For an even more general form of $\Gamma$ we mention two points. First, all configurations with $\partial_{\hat\rho}\xi=0~,~\xi\neq 0~,~\hat R_{\hat\mu\hat\nu\hat\rho\hat\sigma}=0$ are always solutions of the $d$-dimensional field equations derived from the action \eqref{AA1a}. This follows from the simple observation that $F$ must involve at least two powers of $\hat R_{\hat\mu\hat\nu\hat\rho\hat\sigma}$ or its covariant derivatives. Therefore $F$ does not contribute to the extremum condition for $\Gamma$ if 
$\hat R_{\hat\mu\hat\nu\hat\rho\hat\sigma}=0$. Then $\hat R_{\hat\mu\hat\nu\hat\rho\hat\sigma}=0~,~\xi=\xi_0~,~\Lambda=0$ is indeed a solution, as stated in (i) (ii). For the rather wide conditions in presence of a non-polynomial $F(\hat R_{\hat\mu\hat\nu\hat\rho\hat\sigma})$ stated in the first paragraph the essential point is that $\hat R_{\hat\mu\hat\nu\hat\rho\hat\sigma}=0$ remains a solution, and $F(\hat R_{\hat\mu\hat\nu\hat\rho\hat\sigma}=0)=0$. 

For the solutions with $\hat R_{\hat\mu\hat\nu\hat\rho\hat\sigma}=0$ the effective four dimensional action \eqref{AA4a} has $V=0$ and an effective squared Planck mass
\begin{equation}\label{AA8a}
\chi^2=\int_y(g^{(D)})^{1/2}\sigma\xi^2,
\end{equation}
with $\int_y$ an integral over the $D$ internal coordinates $y^\alpha~,~g^{(D)}$ the determinant of the internal metric $g^{(D)}_{\alpha\beta}$ and $\sigma$ the warping factor according to eq. \eqref{2c}. For the example of a $D$-dimensional torus with finite volume $\Omega_D$, and $\sigma=1$, one finds a finite squared Planck mass $\chi^2=\Omega_D\xi^2$. The flat torus solution has an isometry of internal space $U(1)^D$. The corresponding four-dimensional gauge theory has finite non-zero gauge couplings if $\Omega_D$ is finite. This proves at least the existence of asymptotic solutions with the desired general properties. On the other hand, $d$-dimensional Minkowski space with infinite $\Omega_D$ or other flat geometries with infinite $\Omega_D$, are also solutions, but do not lead to an acceptable four-dimensional description. 

The existence of the flat torus solutions leads to an important statement: The most general form of a dilatation symmetric effective action in the absence of a potential for $\xi$ admits always solutions in the flat phase, for which the effective four-dimensional constant $\Lambda$ vanishes, while four-dimensional gravity with $\chi^2>0$ is well behaved. This holds independently of all possible couplings appearing in $\Gamma$. Quantum  fluctuations influence the values of the effective couplings in $\Gamma$. As long as they do not induce a non-polynomial dilatation symmetric potential $V(\xi)$ there are always solutions with $\hat R\hmnrs=0$ and $\Lambda=0$. In this sense the flat phase is stable with respect to quantum fluctuations. The dilatation symmetric fixed point is often reached asymptotically for $\xi\to\infty$. In this case of is sufficient that $V(\xi)$ plays no role in this limit - we will discuss examples of this type in sect. \ref{cosmologicalrunaway}. Having established the existence of solutions in the flat phase, the only remaining issue remains if the class of solutions in the flat phase also comprises realistic particle physics realizations, for example with non-abelian gauge symmetries arising from non-abelian isometries of the internal metric. 

Our second point states that $F$ must not necessarily have the most general possible form. At this place we recall that the effective action \eqref{AA1a} is only supposed to be the part of the action which determines the asymptotic solution in case of a runaway towards a fixed point with restored dilatation symmetry. The contribution of further terms, that violate dilatation symmetry, only vanishes for the fixed point for $t\to\infty$. On the other hand, for this fixed point some of the couplings in $F$ may vanish. For example, if this concerns the parts in $F_H$ that only involve $\hat C_{\hat\mu\hat\nu\hat\rho\hat\sigma}$, solutions with $\xi=const,~\hat R_{\hat\mu\hat\nu}=0$ necessarily lead to a vanishing four-dimensional cosmological constant even for 
$\hat C_{\hat\mu\hat\nu\hat\rho\hat\sigma}\neq 0$.

\medskip\noindent
{\bf 4. General extrema with $\Lambda=0$}

We finally give a simple condition that a dilatation symmetric effective action leads to solutions with $\Lambda=0$. This condition holds for an arbitrary form of $\Gamma$, including a possible non-polynominal potential for $\xi$. Let us define the functional $\bar W$
\be\label{A1}
\bar W = \int_y (g^{(D)} (y))^{1/2} \sigma^2 {\cal L} (y).
\ee
This is analogous to eq. \eqref {2D}, but now evaluated for a metric \eqref{2c} with $g^{(4)}_{\mu\nu}(x)=\eta_{\mu\nu}$ and $\xi=\xi(y)$. The functional $\bar W$ depends then only on the internal metric $g_{\alpha\beta}^{(D)}(y)$ and two $D$-dimensional scalar fields $\sigma(y)$ and $\xi(y)$. Whenever $\bar W$ admits an extremum this corresponds to an extremum of $\Gamma$ with $\Lambda=0$.

By virtue of the scaling \eqref{2F} any extremum of $\bar W$ occurs for $\bar W_0=0$ - the argument is the same as the one given in the introduction. Furthermore, any extremum of $\bar W$ is a solution of the $d$-dimensional field equations. This may be understood most easily by noting that the configurations $\xi=\xi(y)$, 

\be\label{A2}
\hat g\hmn (y)=\left(
\begin{array}{ccc}
\sigma(y)\eta_{\mu\nu}&,&0\\
0&,&g\iD_{\alpha\beta}(y)
\end{array}\right),
\ee
are the most general $d$-dimensional configurations with four-dimensional Poincar\'e-symmetry. An extremum in this subspace is automatically an extremum in the space of arbitrary $d$-dimensional fields. This follows from the classification of $d$-dimensional fields in terms of representations of the Poincar\'e group. The above $D$-dimensional configurations comprise all singlets. Therefore the other representations must appear at least quadratic in $\Gamma$, such that setting them to zero solves the field equations. Of course, one may verify this argument also by an explicit computation of the field equations. Any extremum of $\bar W$ is an extremum of $W(x)$ as defined in eq. \eqref{2D}. As we have argued before, this implies that is an extremum of $\Gamma$. The vanishing of the cosmological constant, $\Lambda=0$, is manifest from the configuration \eqref{A2} and confirms our general discussion.

Finding an extremum of $\bar W$ amounts to a standard problem for a $D$-dimensional euclidean field theory with metric $g_{\alpha\beta}^{D}(y)$ and two scalar fields $\sigma(y),\xi(y)$. The functional $\bar W$ is invariant under general coordinate transformations in $D$-dimensions. With respect to the scaling 
\be\label{A3}
g_{\alpha\beta}^{D}\to \alpha^2 g_{\alpha\beta}^{D}~,~\xi\to\alpha^{-\frac{d-2}2}\xi~,~\sigma\to\sigma
\ee
the functional $\bar W$ has a definite scaling dimension, $\bar W \to\alpha^{-4}\bar W.$ On the other hand, we know that ${\cal L}(y)$ does not involve any coupling with dimension mass or length. We therefore expect the presence of a new $D$-dimensional dilatation symmetry. This is indeed realized, and $\bar W$ remains invariant under the transformation
\be\label{A4}
g_{\alpha\beta}^{D} \to \beta^2 g_{\alpha\beta}^{D}~,~\xi\to \beta^{-\frac{D+2}2} \xi~,~\sigma\to\beta^2\sigma.
\ee
The combination of the transformations \eqref{A3}, \eqref{A4} amounts to the scaling property, with invariant $g_{\alpha\beta}^{D}, \xi$,
\be\label{A5}
\sigma \to\gamma^2 \sigma ~,~ \bar W \to \gamma^4\bar W~,~ {\cal L}\to {\cal L}.
\ee
This accounts for the particular $\sigma$-dependence of $\bar W$ in eq. \eqref{A1} and implies that ${\cal L}$ involves only derivative terms  $\sim\partial_y\ln \sigma$.

The existence of extrema of $\bar W$ seems rather generic and we conclude that the ``flat phase'' of extrema of $\Gamma$ with $\Lambda=0$ is not empty. The flat phase precisely consists of all extrema of $\bar W$. This holds since extrema of $\Gamma$ with maximal four-dimensional symmetry and $\Lambda=0$ are precisely the configurations \eqref{A2}. For a suitable form of ${\cal L}$ the flat phase may comprise geometries with an interesting particle physics, for example spaces with a non-abelian isometry.

\section{Quasistatic solutions}
\label{quasi}
The simple arguments in the preceding section imply that for a large class of dilatation symmetric effective actions a dynamical tuning mechanism for the effective four-dimensional cosmological constant $\Lambda$ must be at work, such that $\Lambda$ vanishes independently of the parameters of the effective action and the details of the solution. Important ingredients of this ``adjustment of the cosmological constant'' can be understood in terms of the effective four-dimensional action \eqref{AA4a}. The general adjustment mechanism will be discussed in sect. \ref{adjustment}.

We are interested in ``quasistatic solutions'' for which four-dimensional space has maximal symmetry (Minkowski or (anti-) de Sitter space), while the internal geometry (including a possible warping) does not depend on the four-dimensional coordinates $x^\mu$. These quasistatic solutions are assumed to be approached for $t\to \infty$. For any finite cosmological time the Universe is still evolving in time. For late time, however, the evolution towards the asymptotic solution becomes slow. The quasistatic solution becomes then a valid approximation for phenomena characterized by not too large length and time scales. This is comparable to Minkowski space being a valid approximation to the present ``cosmological background metric'' on scales much smaller than the horizon. (Of course, the distinction between vanishing and very small $\Lambda$ becomes relevant only for sufficiently large scales.)

In this section we demonstrate that for a wide and generic class of higher dimensional solutions $(\tilde G\geq 0$, see below) all possible quasistatic solutions must have $\Lambda=0$. The essence of this argument are instabilities that exclude any quasistatic solutions with a positive or negative four-dimensional cosmological constant. We further show that possible solutions with $\tilde G<0$, finite $l$ and $\xi\neq 0$ are unstable. Our arguments hold for an arbitrary form of a dilatation symmetric effective action (not necessarily polynomial in the curvature tensor) which depends on $\xi$ only quadratically. In this situation all stable quasistatic solutions with $\xi\neq 0$ and finite characteristic length of internal space belong to the flat phase with $\Lambda=0$. 

\medskip\noindent
{\bf 1. Effective four-dimensional theory}

In the effective four-dimensional theory the effective Planck mass $\chi$ (cf. eqs. \eqref{AA4a}, \eqref{AA8a}) can depend on the values of four-dimensional scalar fields. We may choose a field basis where $\chi(x)$ itself is interpreted as a scalar field. For quasistatic solutions we can neglect in $\Gamma^{(4)}$ all derivative terms for  $\chi$ as well as for other scalar fields. One could restrict the discussion to $|\Lambda|\ll\chi^2$, such that we can also neglect all invariants involving higher powers of the four-dimensional curvature tensor. The effective Planck mass $\chi$ depends on the value of $\xi$ and the characteristic length scale $l$ of internal space
\begin{equation}\label{A1a}
\chi^2=l^D\bar\xi^2-2\tilde G l^{-2},
\end{equation}
with $D$ the dimension of internal space, $D=d-4$. 

More precisely, we define $l$ by the relation
\be\label{16A}
\int_y(g^{(D)})^{1/2}\sigma^2=l^D.
\ee
A suitable average of $\xi$ over internal space is denoted by $\bar \xi$, where the normalization of $\bar \xi$ is fixed such that the coefficient of the first term in eq. \eqref{A1a} equals one,
\be\label{16B}
\int_y(g^{(D)})^{1/2}\sigma^2\xi^2=l^D\bar\xi^2.
\ee
The term $\sim \tilde G$ arises from an expansion of $F$ linear in the four dimensional curvature scalar $R^{(4)}$ and integrated over internal space. If we denote $F=GR^{(4)}+\dots$ the coefficient $\tilde G$ reads
\be\label{16C}
\tilde G=l^2\int_y(g^{(D)})^{1/2}\sigma G.
\ee

Similarly, the effective potential $V$ depends on $\bar\xi$ and $l$ according to
\begin{equation}\label{A2a}
V=\tilde Q\bar\xi^2l^{D-2}+\tilde F l^{-4},
\end{equation}
with $\tilde F$ the appropriate dimensionless integral over internal space of $F$, evaluated for $g^{(4)}_{\mu\nu}=\eta_{\mu\nu}$, while $\tilde Q$ arises from the corresponding integral of the first two terms in eq. \eqref{AA1a}. More explicitely, $\tilde F$ is defined by
\be\label{17A}
\tilde F=l^4\int_y(g^{(D)})^{1/2}\sigma^2 
F(R^{(int)}\hmnrs),
\ee
with $R^{(int)}\hmnrs$ the ``internal part'' of the higher-dimensional curvature tensor which is found for a solution of the type given by eq. \eqref{2c} by replacing $g^{(4)}_{\mu\nu}(x)\to\eta_{\mu\nu}$. The coefficient $\tilde Q$ obtains as 
\be\label{17B}
\tilde Q=\frac12\bar\xi^{-2}l^{2-D}\int_y(g^{(D)})^{1/2}\sigma^2
(\zeta\partial^\alpha\xi\partial_\alpha\xi-\xi^2R^{(int)}).
\ee

We will investigate extrema of $\Gamma^{(4)}$ with respect to free four-dimensional fields $g^{(4)}_{\mu\nu}(x),l(x)$ and $\bar\xi(x)$. For any arbitrary higher dimensional solution with $x$-independent $\xi(y)$ we may multiply $\xi(y)$ by a free $x$-dependent factor, turning $\bar\xi$ in eq. \eqref{16B} to a four-dimensional scalar field. A similar procedure for $x$-dependent factors multiplying $g^{(D)}_{\alpha\beta}(y)$ or $\sigma(y)$ promotes $l(x)$ in eq. \eqref{16A} to a four-dimensional scalar field. Since $\chi(x)$ can be expressed by eq. \eqref{A1a} in terms of $\bar\xi(x)$ and $l(x)$, we do not need to consider it as an independent scalar field. We note that the definitions of $\tilde G,\tilde F, \tilde Q$ are not affected by such $x$-dependent factors. 

We assume that all other fields expect $\bar \xi,l$ and $g^{(4)}_{\mu\nu}$ are taken at values which correspond to solutions of their respective field equations. We can therefore discuss the characteristic behavior of possible solutions of the field equations derived from the effective four-dimensional action $\Gamma^{(4)}$ by variation with respect to $g^{(4)}_{\mu\nu},l$ and $\bar\xi$. They depend on the values of three dimensionless constants $\tilde G,\tilde F, \tilde Q$. We observe that the generic form of eqs. \eqref{A1a}, \eqref{A2a} purely follows from dimensional analysis in case of dilatation symmetry, under the assumption that the $\xi$-dependence of $\Gamma$ is quadratic. The combination of eqs. \eqref{AA4a}, \eqref{A1a}, \eqref{A2a} contains all essential ingredients of our analysis. From the analysis in sect. \ref{dilatation} we learn
that the values of $\tilde G,\tilde F,\tilde Q$ are not arbitrary - often particular values as $\tilde F=\tilde Q=0$ are singled out, consistent with the extremum condition for $\Gamma$ with respect to variations of the degrees of freedom not appearing explicitly in eqs. \eqref{A1a}, \eqref{A2a}. This extends to a large class of warped solutions which will be discussed in the accompanying paper \cite{P2}.

We are interested in maximally symmetric spaces obeying
\begin{equation}\label{A3a}
R^{(4)}=4\Lambda=\frac{4V}{\chi^2},
\end{equation}
with $\Lambda$ the four-dimensional cosmological constant. Insertion into eqs. \eqref{AA4a}, \eqref{2E} yields
\begin{equation}\label{A4a}
\Gamma^{(4)}=\int_x(g^{(4)})^{1/2}W~,~W=V-2\Lambda\chi^2.
\end{equation}
We investigate possible solutions of the higher dimensional field equations or, equivalently, the four-dimensional field equations, with time- and space independent $\bar\xi$ and $l$. They must correspond to an extremum of $W(\bar\xi,l)$, evaluated for fixed $\Lambda$. The problem is then reduced to the discussion of extrema of a simple function of two variables $\bar\xi$ and $l$.

An extremum of $W(\bar\xi,l)$ is not necessarily an extremum of $W(x)$, eq. \eqref{2D}, with respect to variations of $g^{(4)}_{\mu\nu}$. One may therefore find solutions belonging either to the flat or non-flat phase. We will find that for a large part of the space of parameters ($\tilde F,\tilde Q,\tilde G$) no quasistatic solutions with $\Lambda\neq 0$ exist. This concerns the range $\tilde G\geq 0$, where $W$ becomes unstable whenever $\Lambda\neq 0$. Only solutions with $\Lambda=0$ remain possible as candidates for asymptotic quasistatic solutions. In other words, $\tilde G\geq 0$ forces the solutions to belong to the flat phase. For $\tilde G<0$ both the flat and the non-flat phase is possible. Solutions with $\Lambda\neq 0$ are found to be unstable, however, if $\bar\xi\neq 0$ such that all stable quasistatic solutions belong to the flat phase or have $\bar\xi=0$.  

\medskip\noindent
{\bf 2. Classification of possible quasistatic solutions}

The quasistatic solutions correspond to extrema of $W$ with respect to variations of $\bar\xi$ and $l$, where $\tilde F, \tilde G, \tilde Q$ and $\Lambda$ are considered as fixed parameters. It is convenient to use scalar fields with canonical dimension,
\begin{equation}\label{A5a}
\varphi_\xi=\bar\xi l^{\frac{D}{2}}~,~\varphi_l=l^{-1},
\end{equation}
such that
\begin{equation}\label{8a}
V=\tilde Q\varphi^2_\xi\varphi^2_l+\tilde F\varphi^4_l~,~\chi^2=\varphi^2_\xi-2\tilde G\varphi^2_l,
\end{equation}
and
\begin{equation}\label{A6a}
W=\tilde Q\varphi^2_\xi\varphi^2_l+\tilde F\varphi^4_l-2\Lambda \varphi^2_\xi+4\tilde G\Lambda\varphi^2_l.
\end{equation}
The discussion reduces then to an investigation of polynomials with quartic and quadratic terms. It is easy to verify that for all extrema of $W$ the condition \eqref{A3a}, $V-\Lambda\chi^2=W+\Lambda\chi^2=0$, or 
\begin{equation}\label{FAF}
\tilde Q\varphi^2_\xi\varphi^2_l+\tilde F\varphi^4_l-\Lambda\varphi^2_\xi+2\tilde G\Lambda\varphi^2_l=0
\end{equation}
is obeyed.

\begin{figure}[h!tb]
\centering
\includegraphics[scale=0.36]{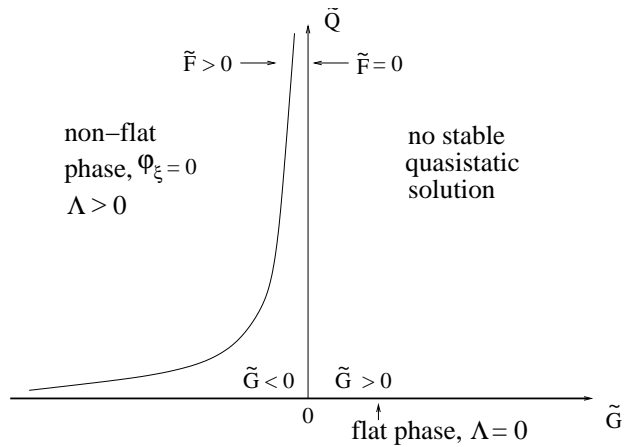}
\caption{Stable extrema of the effective action for finite compactification scale $l$.}
\label{fig1}
\end{figure}

We first consider the case $\tilde F=\tilde G=0$. For example, this is realized for $F=0$ in eq. \eqref{AA1a}. We are interested in solutions with $\varphi_\xi\neq 0$. The extremum condition for $\varphi_l$ reads 
\begin{equation}\label{A7a}
\tilde Q\varphi_l=0,
\end{equation}
such that either $\tilde Q=0$ or $\varphi_l=0$. On the other hand, the extremum condition for $\varphi_\xi$ implies for $\varphi_\xi\neq 0$ 
\begin{equation}\label{A8a}
\tilde Q\varphi^2_l=2\Lambda.
\end{equation}
We conclude from eq. \eqref{A7a} that $\Lambda$ must vanish if $\varphi_\xi\neq 0$. This also follows from eq. \eqref{A3a} since $V$ vanishes if eq. \eqref{A7a} is obeyed. 

Solutions with $\varphi_l=0$ ``solve'' the cosmological constant problem, but also lead to vanishing four-dimensional gauge couplings (as well as other dimensionless couplings like Yukawa couplings if fermions are added). Indeed, an isometry of internal geometry leads to a gauge symmetry in four dimensions, and one finds for the gauge coupling $g$
\begin{equation}\label{A9a}
\frac{1}{g^2}=a_1\bar\xi^2 l^{D+2}+a_2
\end{equation}
where $a_2$ vanishes if the term $\sim F$ is absent. Thus $g$ is nonzero only if
\begin{equation}\label{A10a}
\omega=\bar\xi l^{\frac{D+2}{2}}=\frac{\varphi_\xi}{\varphi_l}
\end{equation}
remains finite. The solution of interest for realistic theories (finite $\omega$) requires $\varphi_l\neq 0$ and therefore $\tilde Q=0$. For $\tilde Q=0,\Lambda=0$ one has $W=V=0$ such that neither $\varphi_l$ nor $\varphi_\xi$ (or $\omega$) are fixed by the extremum condition for $W$. One could indeed find generic solutions with $\tilde Q=0$. 

Next we include possible terms $\sim \tilde F,\tilde G$. For a polynomial effective action they can arise in even dimensions from the higher order curvature invariant $F$. Again, one finds possible solutions with $\varphi_l=0$ for which an extremum at $\varphi_\xi\neq 0$ implies $\Lambda=0$. For realistic solutions with finite $\omega$ and therefore $\varphi_l\neq 0$ the extremum condition for $\varphi_l$ reads
\begin{equation}\label{A11a}
\tilde Q\varphi^2_\xi+2\tilde F\varphi^2_l+4\tilde G\Lambda=0.
\end{equation}
Furthermore, for $\varphi_\xi\neq 0$ eq. \eqref{A8a} must hold, implying
\begin{equation}\label{A12a}
\tilde Q(\omega^2+2\tilde G)+2\tilde F=0.
\end{equation}
The coefficient $\tilde F$ should be positive definite, $\tilde F\geq 0$, in order to avoid an instability of the model - otherwise $V$ would go to $-\infty$ for $\varphi_l\to\infty$. This is guaranteed if $F$ in eq. \eqref{AA1a} has the necessary positivity properties. Stability also requires $\tilde Q\geq 0$. For a large class of generic higher dimensional solutions one further finds $\tilde G\geq 0$. In consequence, any solution of eq. \eqref{A12a} requires
\begin{equation}\label{A13a}
\tilde Q=\tilde F=0.
\end{equation}
In turn, we infer $\Lambda=0$ from eq. \eqref{A8a} or from $V=0$ and eq. \eqref{A3a}. 

In principle, there can be other types of solutions, as $\varphi_l\neq 0,\varphi_\xi=0$. For such solutions the first two terms in eq. \eqref{AA1a} may be omitted. The extremum condition for $\varphi_l$, $\tilde F\varphi^2_l+2\tilde G\Lambda=0$, fixes
\begin{equation}\label{A14a}
\Lambda=-\frac{\tilde F\varphi^2_l}{2\tilde G}.
\end{equation}
Such solutions would lead to stable four-dimensional gravity only for $\tilde G<0$, cf. eq. \eqref{A1a}. For this class of situations a positive nonzero $\Lambda$ is possible. More generally, for negative $\tilde G<-\tilde F/\tilde Q$ eq. \eqref{A12a} has also solutions for $\tilde F\geq 0,\tilde Q>0$, with $\omega^2>0$, obeying
\begin{equation}\label{A15a}
\Lambda=\frac{\tilde Q\varphi^2_l}{2}=-\frac{\tilde F\varphi^2_l}{\omega^2+2\tilde G}\geq 0.
\end{equation}
We conclude that only two types of solutions exist with $\varphi_l\neq 0$ and stable gravity:
\begin{eqnarray}\label{A16a}
\textup{(i)}&&\tilde F=\tilde Q=0~\Rightarrow~ \Lambda=0~,~\omega^2>2\tilde G\nonumber\\
\textup{(ii)}&&\tilde F\geq 0~,~\tilde Q>0~,~\tilde G<0~\Rightarrow~\Lambda>0.
\end{eqnarray}
The first type (i) belongs to the flat phase whereas the second type (ii) characterizes the non-flat phase. One could concentrate on $\tilde G\geq 0$ where only the type (i) is possible. All such solutions lead to a vanishing cosmological constant. 

For arbitrary solutions with $\varphi_l\neq 0$ we note that $\tilde Q=0$ requires $\varphi_\xi=0$ or $\Lambda=0$. For $\Lambda=0,\varphi_\xi\neq 0$ also $\tilde F=0$ is needed. Similarly, for $\tilde F=0$ and $\tilde G\geq 0$ one always needs $\tilde G\Lambda=0$ and $\tilde Q\varphi^2_\xi=0$, which in turn requires $\tilde Q=0$ or $\varphi_\xi=0$. For $\varphi_\xi\neq 0$ this requires $\tilde Q=0,\Lambda=0$. For $\varphi_\xi=0$ we may have the trivial solution $\tilde F=\tilde G=0$, with $\Lambda$ and $\varphi_l$ undetermined that we have omitted in eq. \eqref{A16a}.

Finally, there exists always a solution $\varphi_\xi=\varphi_l=0$. For $\tilde G>0$ it corresponds to an unstable saddlepoint if $\Lambda\neq 0$, while for $\Lambda=0$ it is stable. From eq. \eqref{A6a} one concludes $\chi^2=0$ and one may worry that this leads to a four dimensional theory without dynamical gravity. The issue is, however, more complicated since we should consider the limiting process $\varphi_\xi\to 0,\varphi_l\to 0$. We first note that $\omega$ may be finite in this limiting process such that non-vanishing gauge couplings are possible. Typical particle masses in the effective four-dimensional theory have contributions $\sim \varphi_l$ or $\sim \omega\varphi_l$ and we take for simplicity
\begin{equation}\label{A17}
m^2_p=b_1\varphi^2_l+b_2\omega^2\varphi^2_l.
\end{equation}
What counts for observation are ratios between particle masses and the Planck mass
\begin{equation}\label{A18}
\frac{m^2_p}{\chi^2}=
\frac{b_1\varphi^2_l+b_2\omega^2\varphi^2_l}{\omega^2\varphi^2_l-2\tilde G\varphi^2_l}.
\end{equation}
They remain finite in the limit $\varphi_l\to 0$. For cosmology the ratio $\Lambda/\chi^2\sim\Lambda/\varphi^2_l$ is relevant. For $\Lambda\neq 0$ this quantity diverges for $\varphi_l\to 0$. However, for $\tilde G>0,\Lambda\neq 0$ the unstable point $\varphi_\xi=\varphi_l=0$ is also not approached for asymptotic solutions. In contrast, for $\Lambda=0$ we find a perfectly acceptable limit $\varphi_l\to 0,\varphi_\xi\to 0$, such that the generic solution $\varphi_l=\varphi_\xi=0$ remains interesting. In the vicinity of this solution the quartic terms in $W\sim \tilde F,\tilde Q$ become irrelevant. The approach to $\varphi_\xi=\varphi_l=0$ therefore realizes effectively $\tilde F=\tilde Q=0$ and induces solutions similar to the type (i) in eq. \eqref{A16a}. 

\medskip\noindent
{\bf 3. Phases of solutions and parameter variations}

Our discussion reveals that the dilatation symmetric potential $W$ shows features that are not common for generic forms of effective potentials. Small changes of the parameters $\tilde F$ or $\tilde Q$ or of $\Lambda$ can change qualitatively the behavior of its possible extrema. This can be understood already for the simplest case $\tilde F=\tilde G=0$ where 
\begin{equation}\label{A19}
W=\tilde Q\varphi^2_\xi\varphi^2_l-2\Lambda\varphi^2_\xi.
\end{equation}
Consider first the variation of $\varphi_l$ at fixed $\varphi_\xi$. For $\tilde Q\varphi^2_\xi>0$ one finds a minimum for $W$ at $\varphi_l=0$, while $W$ becomes unstable for $\tilde Q\varphi^2_\xi<0$. For the boundary case $\tilde Q\varphi^2_\xi=0$ the value of $\varphi_l$ remains undetermined. Variation with respect to $\varphi_\xi$ at fixed $\varphi_l$ leads to a similar situation. A minimum at $\varphi_\xi=0$ occurs if $\tilde Q\varphi^2_l>2\Lambda$, and the opposite case $\tilde Q\varphi^2_l<2\Lambda$ leads to instability. For the boundary case $\tilde Q\varphi^2_l=2\Lambda$ the value of $\varphi_\xi$ remains undetermined. All possible extrema of $W$ belong to the boundary case
\begin{equation}\label{A20}
\tilde Q\varphi^2_\xi=0~,~\tilde Q\varphi^2_l=2\Lambda.
\end{equation}
In turn, the only possible solution with nonvanishing $\varphi_l$ and $\varphi_\xi$ requires $\tilde Q=0,\Lambda=0$. Both $\varphi_l$ and $\varphi_\xi$ remain undetermined for this case. An arbitrarily small nonvanishing value of $\tilde Q$ changes the solution qualitatively, either towards instability or a vanishing value for $\varphi_\xi$. 

For the more general case \eqref{A16a} the only solution with $\varphi_\xi\neq 0$ corresponds again to the boundary case between instability and $\varphi_\xi=0$, which is realized if the condition \eqref{A8a} is met. We have seen that all solutions with $\varphi_l\neq 0$ require $\tilde F=0$. We then replace the condition \eqref{A20} with 
\begin{equation}\label{A21}
\tilde Q\varphi^2_\xi=-4\tilde G\Lambda~,~\tilde Q\varphi^2_l=2\Lambda.
\end{equation}
For $\tilde G\geq 0$ the only solution occurs for $\tilde Q=0~,~\Lambda=0$. 

The sensitivity of the existence of the flat or non-flat phase to the precise choice of the parameters $\tilde Q,\tilde F$ is also visible in Fig. \ref{fig1}. At first thought one may be tempted to treat parameters as $\tilde F$ and $\tilde Q$ as generic parameters that could assume any values. In this view the choice $\tilde Q=\tilde F=0$ may look like an artificial tuning of parameters and one would conclude that for generic parameters with $\tilde G>0$ no quasistatic solution exists. However, the parameters $\tilde F, \tilde Q$ are not arbitrary. They result from possible solutions of the higher dimensional field equations. Due to their dependence on infinitely many effective four-dimensional fields they may be restricted to certain values if we consider extrema of $\Gamma$ with respect to all these fields. We will see that this is precisely what happens. We have seen in the preceding section that solutions belonging to the flat phase exist generically. Therefore for $\tilde G\geq 0$ the values of $\tilde F$ and $\tilde Q$ have to adjust themselves to zero in order to be consistent with these solutions. We will discuss this issue of adjustment in more detail in sect. \ref{adjustment}.

\medskip\noindent
{\bf 4. Stability of possible solutions}

We have not yet discussed the stability of the various solutions.  Stability requires that the matrix of second derivatives of $W$ with respect to $\varphi_l$ and $\varphi_\xi$ (at fixed $\Lambda$) should not have negative eigenvalues. For the solutions (i) in the flat phase we have $\tilde F=\tilde Q=\Lambda=0$ and therefore $W$ vanishes identically, consistent with the stability requirement. For the solutions (ii) in the non-flat phase for $\tilde G<-\tilde F/\tilde Q$ we find for $\tilde m^2_{ij}=\frac12\partial^2 W/\partial\varphi_i\partial\varphi_j~,~(\varphi_1,\varphi_2)=
(\varphi_l,\varphi_\xi)$
\begin{eqnarray}\label{H1}
\tilde m^2&=&\left(\begin{array}{ccc}
\tilde Q\varphi^2_\xi+6\tilde F\varphi^2_l+4\tilde G\Lambda&,&2\tilde Q\varphi_\xi\varphi_l\\
2\tilde Q\varphi_\xi\varphi_l&,&\tilde Q\varphi^2_l-2\Lambda
\end{array}\right)\nonumber\\
&=&\left(\begin{array}{ccc}
4\tilde F&,&2\tilde Q\omega\\
2\tilde Q\omega&,&0
\end{array}\right)\varphi^2_l.
\end{eqnarray}
For the second expression we have used the extremum conditions \eqref{A8a}, \eqref{A11a} for $\varphi_\xi\neq 0,\varphi_l\neq 0$. The corresponding eigenvalues of $\tilde m^2$ are
\be\label{B8}
\tilde m^2_\pm=2(\tilde F\pm \sqrt{\tilde F^2+\tilde Q^2\omega^2}~).
\ee
One eigenvalue $\tilde m^2_-$ is negative for all $\tilde Q\neq 0,\omega\neq 0$. Therefore stability requires $\tilde Q=0$, in contradiction to the requirement $\tilde Q>0$ for the existence of solutions in the non-flat phase. Thus the solutions of the type (ii) are unstable. We arrive at the important conclusion that all stable quasistatic solutions must have $\Lambda=0$ if $\Gamma$ is dilatation symmetric and its $\xi$-dependence is quadratic. 

An overview of the situation may be gained by discussing the behavior of $W$ in the $\tilde F,\tilde Q$ plane $(\tilde F\geq 0,\tilde Q\geq 0)$. (We omit the extrema with $\varphi_l=\varphi_\xi=0$ discussed before.) Consider first the case $\tilde F=0,\tilde Q>0$. Extrema of $W$ where either $\varphi_\xi$ or $\varphi_l$ differ from zero are only possible for $\Lambda\neq 0$, obeying either $\tilde Q\varphi^2_l=2\Lambda$ or $\tilde Q\varphi^2_\xi=-4\tilde G\Lambda$ or both. Therefore at least one of the diagonal entries of the matrix $\tilde m^2$ vanishes and all possible extrema must be unstable. Second we investigate the shape of $W$ for $\tilde F>0$. In this case the partial extrema with respect to $\varphi_l$ either have $\varphi_l=0$ or obey
\begin{equation}\label{H2}
\varphi^2_l=-\frac{\tilde Q\varphi^2_\xi+4\tilde G\Lambda}{2\tilde F}.
\end{equation}

We first discuss $\varphi_l\neq 0$ where $\tilde Q\varphi^2_\xi\geq 0$ requires $4\tilde G\Lambda\leq 0$. One may insert eq. \eqref{H2} into $W$ in order to obtain a function depending only on $\varphi_\xi$
\begin{equation}\label{HH}
W(\varphi_\xi)=-\frac{\tilde Q^2}{4\tilde F}\varphi^4_\xi-2\Lambda
\left(\frac{\tilde Q\tilde G}{\tilde F}+1\right)\varphi^2_\xi,
\end{equation}
where the range of $\varphi^2_\xi$ is resticted by $\varphi^2_l>0$ to $\tilde Q\varphi^2_\xi<4\tilde G\Lambda$. In view of the negative coefficient of the term $\varphi^4_\xi$ for $\tilde Q\neq 0$ one easily concludes that $W(\xi)$ can only have a maximum for $\varphi_\xi\neq 0$, clearly indicating the instability. For $\tilde Q>0,\varphi_\xi=0$ stability requires $\Lambda(\tilde Q\tilde G/\tilde F+1)\leq 0$ whereas $\varphi^2_l>0$ needs $\tilde G\Lambda<0$. Stable gravity $(\chi^2<0)$ is compatible only with $\tilde G<0,\Lambda>0$. Using eq. \eqref{H2}, $2\tilde F\varphi^2_l+4\tilde G\Lambda=0$, the eigenvalues of $\tilde m^2$ read $-8\tilde G\Lambda$ and $-2\Lambda(\tilde Q\tilde G/\tilde F+1)$. They are both positive provided $\tilde Q\tilde G<-\tilde F$. Thus the stable solutions in the non-flat phase have all $\varphi_\xi=0$. They are shown in Fig. \ref{fig1}. 

The other alternative extremum for $\tilde F>0$, namely $\varphi_l=0$, needs $\Lambda=0$ for $\varphi_\xi\neq 0$. This solution is stable for all $\tilde G$ and $\tilde Q\geq 0$. We conclude that stable solutions in the flat phase occur for generic parameters. However, for $\tilde F>0$ they have $\varphi_l=0$ and are therefore not compatible with nonvanishing gauge couplings (unless also $\varphi_\xi=0)$. Thus realistic asymptotic cosmologies need $\tilde F=0$ not in order to have $\Lambda=0$, but rather in order to find nonzero gauge couplings. As a general feature for $\tilde F>0,\tilde Q>0$, stable extrema occur only if either $\varphi_\xi$ or $\varphi_l$ vanish. For $\tilde Q=0$ all extrema  with $\Lambda\neq 0$ are unstable. (This is different if no field $\xi$ is present in $\Gamma$. Then $\tilde G<0,\Lambda>0$ admits a stable extremum with $\varphi^2_l=-2\tilde G\Lambda/\tilde F$.) For $\tilde Q=0,\Lambda=0,\tilde F>0$ the only stable extremum occurs for $\varphi_l=0$. Finally, for $\tilde F=\tilde Q=0,\tilde G\neq 0$ all extrema with $\varphi_l\neq 0$ or $\varphi_\xi\neq 0$ belong to the flat phase with $\Lambda=0$. They are stable since $W$ vanishes identically. If $\tilde F=\tilde Q=\tilde G=0$ one formally also finds a stable extremum in the non-flat phase if $\varphi_\xi=0$ and $\Lambda<0$. This particular case has no acceptable gravity since $\chi^2=0$ and will be discarded.

In conclusion of this overview all stable extrema in the non-flat phase must have $\varphi_\xi=0$. They are allowed only in a restricted range for $(\tilde F,\tilde G,\tilde Q)$, with $\tilde G<0,\tilde Q<0$. In contrast, the stable extrema in the flat phase exist for all allowed $(\tilde F,\tilde G, \tilde Q)$. Furthermore, extrema in the flat phase with $\varphi_l\neq 0$ are possible for $\tilde F=\tilde Q=0$, while $\tilde F>0$ implies $\varphi_l=0$. We show the stable extrema for $\varphi_l\neq 0$ in Fig. \ref{fig1}. All stable extrema of $\Gamma$ with finite $l$ and nonzero $\xi$ must have a vanishing cosmological constant $\Lambda=0$! 

\medskip\noindent
{\bf 5.  Modifications for non-polynomial dilaton 

\hspace{0.2cm}potential}

The absence of a higher dimensional potential for $\xi$ is important for the properties of possible solutions. Assume that a dilatation invariant non-analytic term $\sim|\xi|^{2d/(d-2)}$ would be added to the effective action \eqref{AA1a}. After dimensional reduction this would add to $V$ and $W$ a term
\begin{equation}\label{A22}
\Delta V=\tilde C|\varphi_\xi|^{2d/(d-2)}|\varphi_l|^{2(d-4)/(d-2)}.
\end{equation}
An extremum in the $\varphi_\xi$-direction for $\varphi_\xi\neq 0$ (at fixed $\varphi_l$) would now determine the value of $\varphi_\xi$ by 
\begin{equation}\label{A23}
(\tilde Q\varphi^2_l-2\Lambda)|\varphi_\xi|+\frac{d\tilde C}{d-2}
|\varphi_l|^{\frac{2(d-4)}{d-2}}
|\varphi_\xi|^{\frac{d+2}{d-2}}=0.
\end{equation}
For $\tilde C>0$ this has solutions for $\Lambda>\tilde Q\varphi^2_l/2$. Extrema with $\varphi_l\neq 0~,~\varphi_\xi\neq 0$ require now
\begin{equation}\label{A24}
(\tilde F+\frac{2}{d}\tilde Q\omega^2)\varphi^2_l=-\Lambda\left[2\tilde G+\left(1-\frac{4}{d}\right)\omega^2\right]
\end{equation}
For $\tilde G,\tilde F$ and $\tilde Q$ positive semidefinite eq. \eqref{A24} can be solved only for $\Lambda<0$, such that eqs. \eqref{A23} and \eqref{A24} cannot be solved simultaneously and no extremum with $\varphi_l\neq 0,\varphi_\xi\neq 0$ exists. On the other hand, for $\tilde G<0$ solutions with $\Lambda>0$ can be realized. 

In presence of $\tilde C\neq 0$ the stability condition for $V$ and $W$ is modified, however, and $\tilde Q\geq 0$ is no longer needed. The behavior for large $\varphi_l,\varphi_\xi$ is dominated by the quartic terms
\begin{equation}\label{A25}
V=(\tilde F+\tilde Q\omega^2+\tilde C|\omega|^{\frac{2d}{d-2}})\varphi^4_l=K(|\omega|)\varphi^4_l.
\end{equation}
Stability requires $K(|\omega|)\geq 0$ or
\begin{equation}\label{A26}
\tilde F\geq 0~,~\tilde C\geq 0~,~\tilde Q\geq \tilde Q_{\textup{min}}=-
\frac d2\tilde F\left(\frac{d-2}{2}\frac{\tilde F}{\tilde C}\right)^{-\frac{d-2}{d}}.
\end{equation}
For $\tilde Q=\tilde Q_{\textup{min}}$ the minimum of $V$ occurs for $V_{\textup{min}}=0$ at
\begin{equation}\label{A27}
\omega^2_{\textup{min}}=-\frac d2\frac{\tilde F}{\tilde Q_{\textup{min}}},
\end{equation}
while for $\tilde Q>\tilde Q_{\textup{min}}$ one has $V_{\textup{min}}>0$. For $\tilde Q=\tilde Q_{\textup{min}}$ all possible solutions have $\Lambda=0$. On the other hand, for $\tilde Q>\tilde Q_{\textup{min}}$ possible solutions must have $\Lambda>0$ since $V_{\textup{min}}>0$. For $\tilde G\geq 0$ eq. \eqref{A24} requires for an extremum $\omega^2>-(d/2)(\tilde F/\tilde Q)$. In other words, for $\tilde C>0$ stability admits negative $\tilde Q$, and for $\tilde Q<0$ one has a minimum of $K(|\omega|)$ for finite and non-zero $|\omega|$. If $V$ is positive at the minimum one will find a nonzero cosmological constant $\Lambda>0$. We have not performed a stability analysis for the new types of possible extrema for $\tilde C>0$. 

At first sight the discussion above seems to indicate that in presence of a non-polynomial potential for $\xi$ the generic solution has $\Lambda>0$. This conclusion is premature, however, since it implicitely assumes that the couplings $\tilde F,\tilde Q, \tilde G$ and $\tilde C$ are more or less arbitrary within their allowed ranges. We have seen, however, in the preceeding section \ref{dilatation},4. that extrema in the flat phase exist rather generically for the most general dilatation symmetric effective action. This includes a possible non-polynomial potential for $\xi$. One concludes that the higher dimensional field equations and extremum conditions for $\Gamma$ precisely single out $\tilde Q_{min}$ for all extrema in the flat phase. On the other hand, we have not yet found consistent extrema of $\Gamma$ that lead to $\tilde Q>\tilde Q_{min}$ and therefore $\Lambda>0$. 

We will not consider the non-analytic term in the following. We can then conclude that all possible stable quasistatic extrema with $\varphi_l\neq 0,\varphi_\xi\neq 0$ must have $\Lambda=0$. They are only possible for $\tilde F=0,\tilde Q=0$. The existence of such solutions follows from sect. \ref{dilatation} and we will discuss their properties extensively in the later parts of this paper.

\section{Extended scaling}
\label{extended}

A particularly simple case arises if the term $\sim F$ in eq. \eqref{AA1a} does not contribute to the field equations for the asymptotic fixed point solution. (For polynomial interactions this is always realized for $d$ odd since no polynomial invariant contributing to $F$ exists at all.) From sect. \ref{dilatation} we learn for this case that $\xi=\xi_0~,~\hat R_{\hat\mu\hat\nu}=0$ is a solution of the field equations and leads to $\Lambda=0$. On the other hand, the discussion of sect. \ref{quasi} implies that all possible quasistatic solutions must have $\Lambda=0$, since $\tilde G=0$. 

\medskip\noindent
{\bf 1. Extended scaling and vanishing cosmological 

\hspace{0.2cm}constant}

For $F=0$ the field equations are invariant under an ``extended scale transformation''
\begin{equation}\label{B1a}
\xi\to \beta\xi~,~\hat g_{\hat\mu\hat\nu}\to\beta^2\hat g_{\hat\mu\hat\nu}.
\end{equation}
Indeed, the curvature scalar transforms as $\hat R\to \beta^{-2}\hat R$, such that the effective action scales proportional to $\hat g^{1/2}$, $\Gamma\to\beta^d\Gamma$. The field equations obtain from variation of $\Gamma$
\begin{equation}\label{B2a}
\frac{\delta\Gamma}{\delta\xi}=0~,~\frac{\delta\Gamma}{\delta\hat g^{\hat\mu\hat\nu}}=0.
\end{equation}
They are invariant under the extended scaling since $\Gamma$ and $\beta^d\Gamma$ lead to the same field equations. The extended scale transformations add to the dilatation transformations
\begin{equation}\label{B3a}
\xi\to\alpha^{-\frac{d-2}{2}}\xi~,~\hat g_{\hat\mu\hat\nu}\to\alpha^2\hat g_{\hat\mu\hat\nu},
\end{equation}
under which $\Gamma$ is assumed to be invariant for the asymptotic solution. Extended scaling is realized whenever the effective action is dilatation symmetric and quadratic in $\xi$. Indeed, a rescaling of $\xi\to\eta\xi,\Gamma\to\eta^2\Gamma$, combined with the dilatations \eqref{B3a}, implies the extended scaling \eqref{B1a}.

The extended scale symmetry helps to understand the tuning mechanism which leads to a vanishing cosmological constant for all quasistatic solutions. For any solution of the field equations \eqref{B2a} $\Gamma$ must vanish. This can be shown as follows. Consider a solution $\xi^{(0)},\hat g^{(0)}_{\hat\mu\hat\nu}$ with $\Gamma^{(0)}=\Gamma[\xi^{(0},\hat g^{(0)}_{\hat\mu\hat\nu}]$. A neighboring field configuration, scaled according to eq. \eqref{B1a} with $\beta=1+\delta\beta$, is also a solution, with $\Gamma^{(0)}(\delta\beta)=(1+\delta\beta)^d\Gamma^{(0)}$. Solutions correspond to extrema of $\Gamma$ such that for two infinitesimally close solutions one has $\partial\Gamma^{(0)}(\delta\beta)/\partial\delta\beta|_{\delta\beta=0}=0$. This implies $\Gamma^{(0)}=0$ and results in $\Lambda=0$, as discussed in sect. \ref{dilatation}. 

Next, we may discuss the extended scaling of the effective four dimensional fields \eqref{A5a}
\begin{eqnarray}\label{B4a}
\varphi_l&\to&\beta^{-1}\varphi_l~,~\varphi_\xi\to\beta^{\frac{d-2}{2}}\varphi_\xi,
\\
g^{(4)}_{\mu\nu}&\to&\beta^2g^{(4)}_{\mu\nu}~,~
R^{(4)}\to\beta^{-2}R^{(4)}~,~\Gamma^{(4)}\to\beta^d\Gamma^{(4)}.\nonumber
\end{eqnarray}
While
\begin{equation}\label{B5a}
\Gamma^{(4)}=\int_x(g^{(4)})^{1/2}
(\tilde Q\varphi^2_\xi\varphi^2_l-\frac12
\varphi^2_\xi R^{(4)})
\end{equation}
has the correct scaling properties, a possible potential term $\sim \tilde F\varphi^4_l$ is not compatible with $V\to\beta^DV$ and therefore not allowed. An asymptotic fixed point where $F$ becomes irrelevant in eq. \eqref{AA1a} can therefore be interpreted as a fixed point with extended scaling symmetry. In this sense a term $\sim F$ can be regarded as an ``anomalous term'' with respect to extended scaling, which vanishes for the asymptotic solution. 

We can combine the extended scale transformations \eqref{B1a} with the dilatations \eqref{B3a} such that $\varphi_\xi$ remains invariant. With dilatations acting as $l\to \alpha l$ and 
\begin{equation}\label{B7a}
\varphi_l\to\alpha^{-1}\varphi_l~,~\varphi_\xi\to\alpha^{-1}\varphi_\xi~,~
g^{(4)}_{\mu\nu}\to\alpha^2g^{(4)}_{\mu\nu},
\end{equation}
we choose $\alpha=\beta^{\frac{d-2}{2}}$ for this purpose, such that the combined transformation reads
\begin{equation}\label{B8a}
\varphi_l\to\epsilon^{-1}\varphi_l~,~\varphi_\xi\to\varphi_\xi~,~g^{(4)}_{\mu\nu}
\to\epsilon^2g^{(4)}_{\mu\nu}~,~\Gamma^{(4)}\to\epsilon^2\Gamma^{(4)}.
\end{equation}
Instead of eq. \eqref{B1a} we will take the transformation \eqref{B8a} as our definition of extended scaling in the following. 

\medskip\noindent
{\bf 2.  Internal scaling}

We observe that for $\tilde Q=0$ the action \eqref{B5a} exhibits a further symmetry
\begin{equation}\label{B6a}
\varphi_l=\gamma^{-1}\varphi_l
\end{equation}
while $\varphi_\xi$ and $g^{(4)}_{\hat\mu\hat\nu}$ remain invariant. This means that the volume of internal space decouples from the four dimensional physics. In other words, one can scale $l\to\gamma l$ without changing $\Gamma^{(4)}$. We may call the transformation \eqref{B6a} ``internal scaling''. (The field $\varphi_l$ corresponds of the volume moduli field in string theory.) Combining internal scaling \eqref{B6a} with extended scaling \eqref{B8a} we arrive at a scaling where only $g^{(4)}_{\mu\nu}$ and $\Gamma^{(4)}$ scale $\sim\epsilon^2$. Any solution leading to a nonzero cosmological constant $V_0$ would contribute to $\Gamma$ a term scaling $\sim(g^{(4)})^{1/2}\sim\epsilon^4$, in contradiction to the scaling $\Gamma\sim\epsilon^2$. Thus only $V_0=0,\Lambda=V_0/\varphi^2_\xi=0$ is compatible with the combination of internal and extended scaling. We observe that $\tilde Q=0$ and therefore internal scaling invariance of the asymptotic solution is necessary for the existence of a quasistatic solution. This shows again that in presence of the extended scaling symmetry all quasistatic solutions must have $\Lambda=0$.

\medskip\noindent
{\bf 3.  Goldstone bosons}

We can extend the four-dimensional action \eqref{B5a} by including derivative terms for the fields $\varphi_\xi$ and $\varphi_l$ which have the generic form 
\begin{eqnarray}\label{B9a}
\Gamma^{(4)}_{\textup{kin}}&=&\frac12\int_x(g^{(4)})^{1/2}(c_\xi l^D\partial^\mu
\bar\xi\partial_\mu\bar\xi+c_l\bar\xi^2
l^D\partial^\mu\ln l\partial_\mu\ln l)\nonumber\\
&=&\frac12\int_x(g^{(4)})^{1/2}
\Big[c_\xi\partial^\mu\varphi_\xi\partial_\mu\varphi_\xi+Dc_\xi
\varphi_\xi\partial^\mu
\varphi_\xi\partial_\mu\ln \varphi_l\nonumber\\
&&+\left(\frac{D^2}{4}c_\xi+c_l\right)\varphi^2_\xi\partial^\mu\ln\varphi_l
\partial_\mu\ln\varphi_l\Big].
\end{eqnarray}
One verifies that $\Gamma_{4,\textup{kin}}$ is invariant under dilatations and internal scaling, and covariant with respect to extended scaling \eqref{B8a}. For $\tilde Q=0$ no potential is present and we can identify $\varphi_\xi$ and $\varphi_l$ with the two Goldstone bosons of spontaneously broken dilatation symmetry and internal scaling symmetry that should be present for any solution with $\varphi_\xi\neq 0,\varphi_l\neq 0$. Alternatively, we can write $\Gamma^{(4)}_{\textup{kin}}$ in terms of $\varphi_\xi$ and $\omega=\varphi_\xi/\varphi_l$,
\begin{eqnarray}\label{B10a}
&&\Gamma^{(4)}_{\textup{kin}}=\frac12\int_x(g^{(4)})^{1/2}
(Z_\xi\partial^\mu\varphi_\xi\partial_\mu\varphi_\xi\nonumber\\
&&+Z_{\xi\omega}\varphi_\xi\partial^\mu\varphi_\xi\partial_\mu\ln \omega+Z_\omega\varphi^2_\xi
\partial^\mu\ln \omega\partial_\mu\ln \omega),
\end{eqnarray}
with
\begin{eqnarray}\label{B11a}
Z_\xi&=&\left(1+\frac D2\right)^2 c_\xi+c_l,\nonumber\\
Z_{\xi \omega}&=&-D\left(1+\frac D2\right)c_\xi-2c_l,\nonumber\\
Z_\omega&=&\frac{D^2}{4}c_\xi+c_l.
\end{eqnarray}

\medskip\noindent
{\bf 4.  Extended scaling anomaly}

One may ask if other interactions in the effective four dimensional theory can be consistent with dilatations and extended scaling. As an example, we include spinors $\psi$ and gauge fields $A^z_\mu$ with field strength $F^z_{\mu\nu}$ and covariant derivatives $D_\mu=\partial_\mu-iA^z_\mu T_z$ ($T_z$ are the appropriate generators of the gauge group):
\begin{eqnarray}\label{B12a}
\Gamma^{(4)}&=&\int_x(g^{(4)})^{1/2}
\Big\{-\frac12\varphi^2_\xi R^{(4)}
+\frac{\tilde Q\varphi^4_\xi}{\omega^2}\nonumber\\
&&+\frac12
(Z_\xi\partial^\mu\varphi_\xi\partial_\mu\varphi_\xi+Z_{\xi\omega}
\varphi_\xi\partial^\mu\varphi_\xi
\partial_\mu\ln\omega\nonumber\\
&&+Z_\omega\varphi^2_\xi\partial^\mu \ln\omega~\partial_\mu\ln\omega)
+i\bar\psi\gamma^\mu D_\mu\psi
\nonumber\\
&&
+\frac14\omega^2 F^{\mu\nu}_zf\left(-{{\cal D}^2}/\varphi^2_\xi~,~\omega^2\right)
F^z_{\mu\nu} 
\Big\}.
\end{eqnarray}
The transformations under extended scaling \eqref{B8a} $(\gamma^\mu=\gamma^me^\mu_m~,~g^{\mu\nu}=e^{\mu}_me^{m\nu})$ are given by 
\begin{equation}\label{B13a}
\psi\to\epsilon^{-1/2}\psi~,~A_\mu\to A_\mu~,~\omega\to\epsilon\omega.
\end{equation}
We have introduced $f$ in order to account for the running of the effective four dimensional gauge coupling as a functional of momentum $(q^2\widehat{=}-{\cal D}^2)$
\begin{equation}\label{B14a}
g^2(q^2)=\rho^{-2}f^{-1}\left(\frac{q^2}{\varphi^2_\xi},\omega^2\right).
\end{equation}
Only the combination ${\cal D}^2\omega^2/\varphi^2_\xi~,~{\cal D}^2=D^\mu D_\mu$, is invariant under dilatations and extended scaling such that extended scaling is preserved only if $f$ depends only on the combination $q^2\omega^2/\varphi^2_\xi=q^2/\varphi_l^2$. However, for low enough momenta the function $f$ is determined by the perturbative four-dimensional beta-function for the running gauge coupling, $\partial g^{-2}/\partial \ln q^2=b~(b>0$ for asymptotically free theories) as
\begin{equation}\label{B15a}
f=\omega^{-2}b \ln(q^2/\tilde \mu^2)+f_0~,~g^{-2}(q^2=\tilde \mu^2)=f_0\omega^2.
\end{equation}
There is no choice of $\tilde \mu$ for which $f$ only depends on $q^2\omega^2/\varphi^2_\xi$. We conclude that the running of four dimensional couplings induces an anomaly for extended scaling. For $\tilde \mu^2=\varphi^2_l=\varphi^2_\xi/\omega^2$ the dilatation symmetry is preserved.

\section{Adjustment of the cosmological constant}
\label{adjustment}

From a four-dimensional point of view it seems that some type of ``adjustment'' or ``tuning'' of the cosmological constant takes place. One may wonder if this is not ``unnatural'' in the sense that particular parameters of the more fundamental higher dimensional theory must be chosen for this tuning of $\Lambda$ to the value zero to happen. We argue in this section that this is not the case. A natural self-tuning mechanism assures $\Lambda=0$ independently of the precise values of parameters in the dilatation symmetric effective action $\Gamma$. Since $\Gamma$ includes all quantum fluctuations, this means that $\Lambda=0$ is realized in the presence of quantum fluctuations. While quantum fluctuations of various fields give nonvanishing individual contributions to $\Lambda$, the sum of all contributions to $\Lambda$ vanishes due to dilatation symmetry at the fixed point. The general arguments why in the presence of symmetries the size of individual contributions can not be taken as an indication for the natural size of the full quantity can be found in ref. \cite{CWCC}.

The essence of the self-tuning mechanism which is at work in our case relies both on dilatation symmetry and the presence of higher dimensions. It can be understood from the discussion of the quantity $W$ in the introduction. Dilatation symmetry assures that any extremum of $W$ has to occur for $W_0=0$ and therefore $\Lambda=0$. In turn, the higher dimensional setting provides the flexibility that an extremum of $W$ exists independently of the precise values of couplings in $\Gamma$. This holds since $W$ is a functional of higher dimensional fields and for very generic situations extrema of functionals exist. In a four-dimensional language, the presence of an infinity of four-dimensional fields overcomes the difficulties of a self-tuning mechanism for a finite number of fields.

As an example, we discuss the simplest situation where the contribution of the higher curvature terms $\sim F$ to the asymptotic solution of the field equations can be neglected. The tuning concerns then the effective constant $\tilde Q$ in the four dimensional potential \eqref{A2a}. As we have seen, for $\tilde G\geq 0$ quasistatic solutions exist only for $\tilde Q=0$, and they lead to a vanishing cosmological constant, $\Lambda=0$.

In general, $\tilde Q$ depends on the geometry of internal space and a possible warping, as well as on the configuration of $\hat\xi$. We have assumed that $\tilde Q$ is evaluated for all fields except $g^{(4)}_{\mu\nu},l$ and $\bar\xi$ taking values corresponding to partial extrema of the action. In other words, it is given by a solution of the field equations for these other fields. Let us investigate this issue in some more detail and consider first the case of one particular degree of freedom $\alpha$, reflecting some particular variation of internal geometry, warping or $\hat\xi$. Then the extremum condition for $\alpha$ reads
\begin{equation}\label{WW1}
\frac{\partial\tilde Q(\alpha)}{\partial \alpha}_{|\alpha_0}=0,
\end{equation}
while the existence of a quasistatic solution requires
\begin{equation}\label{WW2}
\tilde Q(\alpha_0)=0.
\end{equation}
The coincidence of the extremum with a zero of $\tilde Q(\alpha)$ requires a specific form of $\tilde Q$, and this is understood by tuning or adjustment. If we make an arbitrary small change of the function $\tilde Q(\alpha)$, the two conditions \eqref{WW1} and \eqref{WW2} are no longer obeyed simultaneously.

The situation looks more complicated if we include infinitely many degrees of freedom, as appropriate for a higher dimensional theory. Assume that $\tilde Q$ is a functional of $\alpha(y)$, where $y$ are coordinates of internal space 
\begin{equation}\label{WW3}
\tilde Q=\tilde R\big[\alpha(y)\big].
\end{equation}
The extremum condition for $\alpha$ appears now in the form of the field equations
\begin{equation}\label{WW4}
\frac{\delta\tilde R}{\delta\alpha(y)}=0.
\end{equation}
If $\tilde R$ contains derivative terms, as in our case, the most general local solution of the field equation $\alpha_0(y)$ will depend on free integration constants $\gamma_i$, i.e. $\alpha_0(y;\gamma_i)$. It seems now more plausible that an appropriate choice of the integration constants leads to 
\begin{equation}\label{WW5}
\tilde R\big[\alpha_0(y;\gamma_i)\big]=0,
\end{equation}
such that quasistatic solutions with $\tilde Q=0$ become possible. 

The existence of solutions in the flat phase, that we have established in sect. \ref{dilatation} in the absence of a higher dimensional dilaton potential for the most general dilatation symmetric effective action, shows that suitable integration constants $\gamma_i$ can indeed always be found. It is an interesting question to understand ``how large'' is the space of solutions in the flat phase, or what is the space of integration constants $\gamma_i$ consistent with eq. \eqref{WW5}. In the accompanying paper \cite{P2} we discuss in detail spaces with warping, where the integration constants $\gamma_i$ appear directly as integration constants of specific higher dimensional solutions. There we find explicitly that suitable values of the integration constants $\gamma_i$ can indeed be found. 

The existence of integration constants $\gamma_i$ obeying eq. \eqref{WW5} holds for an arbitrary choice of the couplings of the higher-dimensional theory. The ``tuning'' $\Lambda=0$ therefore requires no special choice of parameters. What remains to be investigated are two questions: if the choice of $\gamma_i$ which leads to $\Lambda=0$ corresponds to an acceptable extremum of the effective action, and if other choices with $\Lambda\neq 0$ are equally acceptable. We find in ref. \cite{P2} that $\Lambda=0$ is indeed singled out as an acceptable extremum, a result which is in close connection to the observation that (for $\tilde G\geq 0)$ quasistatic solutions only exist in this case.

This generalizes to a very generic form of the dilatation symmetric effective action and generic solutions. It is now sufficient that an extremum of the functional $W[\alpha(y)]$ exists in order to assure $\Lambda=0$. As we have discussed in the introduction extrema of $W$ correspond to extrema of $\Gamma$.

\section{Dilatation symmetry with higher order curvature terms}
\label{higherorder}

In this section we discuss the role of higher order curvature invariants for the properties of extrema of the dilatation symmetric effective action. We thus consider the general form eq. \eqref{AA1a}, and first assume that $F$ is a polynomial of the curvature tensor and its derivatives. We will later abandon the restriction to polynomial $F$. As discussed in sect. II, the most general form of $F$ is then a polynomial of the irreducible tensors $\hat R,\hat H_{\hat\mu\hat\nu}$ and $\hat C_{\hat\mu\hat\nu\hat\rho\hat\sigma}$ and their covariant derivatives. Indices are contracted with the metric - we omit parity violating contractions with the $\epsilon$-tensor.

\medskip\noindent
{\bf 1. Field equations and condition for $\Lambda=0$}

The field equations for the higher dimensional metric derived from eq. \eqref{AA1a} can be written in the form
\begin{equation}\label{NA}
\xi^2(\hat R_{\hat\mu\hat\nu}-\frac12\hat R\hat g_{\mu\nu})=T^{(\xi)}_{\hat\mu\hat\nu}+T^{(F)}_{\hat\mu\hat\nu},
\end{equation}
while the field equation \eqref{AA3a} for $\xi$ is not affected by the $F$-term. The ``scalar part'' $T^{(\xi)}_{\hat\mu\hat\nu}$ of the energy momentum tensor is given by \begin{eqnarray}\label{85A}
T^{(\xi)}_{\hat\mu\hat\nu}
=\zeta\partial_{\hat \mu}\xi\partial_{\hat\nu}\xi-\frac\zeta 2\partial^{\hat\rho}\xi\partial_{\hat\rho}\xi~\hat g_{\hat\mu\hat\nu}
+D_{\hat\nu}D_{\hat \mu}\xi^2-\hat D^2\xi^2\hat g_{\hat\mu\hat\nu},
\end{eqnarray}
while the ``curvature part'' $T^{(F)}_{\hat\mu\hat\nu}$ obtains from the variation of the $F$-term 
\begin{equation}\label{9}
T^{(F)}_{\hat\mu\hat\nu}=2\frac{\delta F}{\delta\hat g^{\hat\mu\hat\nu}}- F\hat g_{\hat\mu\hat\nu}.
\end{equation}

The scalar field equation implies for constant non-zero $\xi_0$ that the curvature scalar must vanish, $\hat R=0$. In turn, the contraction of the gravitational field equation with $\hat g^{\hat\mu\hat\nu}$ implies $\hat T^{(F)}_{\hat\mu\hat\nu}\hat g^{\hat\mu\hat\nu}=0$. What remains is the field equation fore the traceless part
\begin{equation}\label{10}
\xi^2_0\hat H_{\hat\mu\hat\nu}=2\frac{\delta F}{\delta\hat g^{\hat\mu\hat\nu}}-Fg_{\hat\mu\hat\nu}.
\end{equation}
We may write
\begin{equation}\label{11}
\frac{\delta F}{\delta\hat g^{\hat\mu\hat\nu}}=l^{-d}_0A_{\hat\mu\hat\nu}+\frac12 F\hat g_{\hat\mu\hat\nu}~,~\hat g^{\hat\mu\hat\nu}A_{\hat\mu\hat\nu}=0,
\end{equation}
If $A_{\hat\mu\hat\nu}=0$ one finds $\hat H_{\hat\mu\hat\nu}=0$ and the solution reads \begin{equation}\label{70AA}
\hat R_{\hat\mu\hat\nu}=0~,~\xi=\xi_0.
\end{equation}
On the other hand, for a non-vanishing traceless part of the energy momentum tensor, $A_{\hat\mu\hat\nu}\neq 0$, one also finds a nonvanishing $\hat H_{\hat\mu\hat\nu}=2\xi^{-2}_0l^{-d}_0A_{\hat\mu\hat\nu}$. With $\hat H_{\hat\mu\hat\nu}\sim l^{-2}_0$ one may relate the scales $\xi_0$ and $l_0$ in this case, $l_0\sim\xi^{-\frac{2}{d-2}}_0$. 

For all extrema of the effective action which define a four dimensional geometry according to eq. \eqref{2c}, the value of $\Gamma$ at the extremum obeys
\begin{equation}\label{NB}
\Gamma_0=\int_x(g^{(4)})^{1/2}W^{(0)}_F,
\end{equation}
with $W^{(0)}_F$ given by eq. \eqref{2G}
\begin{equation}\label{NBA}
W^{(0)}_F=\int_y(g^{(D)}_0)^{1/2}\sigma^2_0F_0.
\end{equation}
Here $g^{0)}_0,\sigma_0$ and $F_0$ are evaluated for the extremum. In other words, the first two terms in eq. \eqref{AA1a} do not contribute. This follows from the scalar field equation, together with the extremum condition
\begin{equation}\label{NC}
\int_y\partial_{\hat \mu} (\hat g^{1/2}\xi\partial^{\hat\mu}\xi)=0.
\end{equation}
For regular spaces eq. \eqref{NC} is obeyed automatically, while for singular spaces it guarantees that $\Gamma$ is an extremum with respect to variations $\xi\to\xi\big(1+\epsilon(x)\big)$, where $\epsilon$ is local in four dimensional space \cite{CWCON}, \cite{P2}. 

As we have seen, the vanishing of $\Gamma_0$ implies a vanishing cosmological constant for any consistent local four dimensional effective gravity. We thus conclude that $F_0=0$ implies $\Lambda=0$, while for $F_0\neq 0$ also $\Lambda$ will differ from zero. This yields a simple condition for a vanishing four dimensional cosmological constant in case of a dilatation invariant effective action: The value of the invariant $F$, evaluated for an extremum of the effective action, must vanish
\begin{equation}\label{ND}
F_0=F[g^{(0)}_{\alpha\beta},\sigma^{(0)}]=0.
\end{equation}
For consistency, $F_0$ has to be evaluated for flat four-dimensional space, i.e. for a metric
\begin{equation}\label{NE}
\hat g_{\hat\mu\hat\nu}=\left(\begin{array}{ccc}
\sigma\eta_{\mu\nu}&,&0\\
0&,&g_{\alpha\beta}
\end{array}\right),
\end{equation}
where the internal metric $g_{\alpha\beta}$ and the warp factor $\sigma$ depend on the internal coordinates. We emphasize that for the relation $F_0=0\leftrightarrow\Lambda=0$ a polynomial form of $F$ is not necessary. 

\medskip\noindent
{\bf 2. Polynomial action and simple solutions}

In this section we concentrate on a polynomial form of $F$ in terms of the curvature tensor and its derivatives. The resulting solutions with a vanishing cosmological constant may serve as instructive examples. They can also serve as starting points for ``deformations'' of the effective action where $F$ has no longer a polynomial form. If we disregard parity violating  contractions with the $\epsilon$-tensor, no polynomial dilatation invariant $F$ can be constructed in odd dimensions. Thus for $d$ odd and a restriction to polynomial $F$ the case $F=0$ is the only relevant case. For even $d$ the most general polynomial dilatation invariant $F$ can be built from powers of the curvature tensor and its covariant derivatives. More precisely, it must involve $\frac d2-m$ powers of $\hat R _{\hat\mu\hat\nu\hat\rho\hat\sigma}$, with $m$ the number of covariant derivatives which must be even. In principle, there may also be polynomials with one power of $\xi$ and $\frac d4+\frac12-m$ powers of $\hat R_{\hat\mu\hat\nu\hat\rho\hat\sigma}$. This is only possible for $d=2$ mod $4$. We exclude here such invariants by imposing a discrete symmetry $\xi\to-\xi$. We also disregard a total derivative $(\hat D^2)^{(d-2)/2}\hat R$ - therefore $F$ contains at least two powers of the curvature tensor.

As we have argued in sect. \ref{dilatation}, this implies that the direct product solutions with geometry ${\cal M}^4\times{\cal F}^D$, with ${\cal F}^D$ a flat $D$-dimensional space (e.g. a torus) and constant $\xi$, are an extremum of the effective action. For 
\begin{equation}\label{NF}
\hat R_{\hat\mu\hat\nu\hat\rho\hat\sigma}=0~,~\xi=const
\end{equation}
one obviously finds $F_0=0$, and therefore $\Lambda=0$ for the effective four-dimensional theory which obtains if ${\cal F}^D$ has finite volume. The field equations are obeyed for eq. \eqref{NF} since $T^{(F)}_{\hat\mu\hat\nu}$ must be at least linear in $\hat R_{\hat\mu\hat\nu\hat\rho\hat\sigma}$ if $\Gamma$ is at least  quadratic. Furthermore, for a regular space ${\cal F}^D$ there are no additional ``brane constraints'' \cite{CWCON} and the solution of the field equations is sufficient to guarantee that the configuration \eqref{NF} is indeed an extremum. In case of isometries of ${\cal F}^D$ and finite volume $\Omega_D$ the gauge couplings in the effective four-dimensional gauge theory are nonzero and finite. 

As we have mentioned already, this simple finding establishes that a very wide class of higher dimensional dilatation symmetric effective actions has extrema for which $\Lambda=0$ in the effective four dimensional gravity theory after dimensional reduction. The phase with $\Lambda=0$ (in the language of sect. \ref{quasi}) is not empty. In fact, the only thing needed for the existence of the extremum \eqref{NF} is the absence of a higher-dimensional dilatation symmetric potential $V(\xi)\sim\xi^{2d/(d-2)}$. 

In the presence of higher dimensional dilatation symmetry and the absence of a fractional potential $V(\xi)$ the issue is therefore not the existence of ``compactifications'' with $\Lambda=0$. The interesting question rather concerns an investigation how ``extended'' is the phase with $\Lambda=0$, and if it contains spaces with phenomenologically interesting four-dimensional non-abelian gauge symmetries. 

We next address possible Ricci flat extrema
\begin{equation}\label{NG}
\hat R_{\hat\mu\hat\nu}=0~,~\xi=const.,
\end{equation}
while $\hat R_{\hat\mu\hat\nu\hat\rho\hat\sigma}$ and $\hat C_{\hat\mu\hat\nu\hat\rho\hat\sigma}$ may differ from zero. The configuration \eqref{NG} obeys the field equations if $F$ contains at least two powers of $\hat R_{\hat\mu\hat\nu}$. In this case $T^{(F)}_{\hat\mu\hat\nu}$ is at least linear in $\hat R_{\hat\mu\hat\nu}$ and therefore vanishes by virtue of eq. \eqref{NG}. Then one also finds $F_0=0$ and therefore $\Lambda=0$. Possible extrema are direct product spaces ${\cal M}^4\times {\cal R}^D$, with ${\cal R}^D$ a Ricci flat internal space. (The flat spaces ${\cal F}^D$ are a subclass of ${\cal R}^D$.) We note that the configuration \eqref{NG} is a solution of the field equations for arbitrary warped geometries, not necessarily of the direct product type. For spaces with singularities we have to guarantee, however, that the brane constraints are obeyed \cite{CWCON}. In ref. \cite{P2} we discuss the general case of warped spaces with metric \eqref{2c}, where internal space exhibits the isometry $SO(D)$. In this case no warped solutions with finite $\Omega_D$ are extrema of $\Gamma$, such that only spaces ${\cal M}\times{\cal R}^D$ remain as candidates. It is not known to us if the solution \eqref{NG} contains warped extrema with a smaller isometry group.

We finally observe that any possible extremum of $F$ with respect to variations of the metric must occur for $F=0$. This follows directly from the scaling of $F$ under dilatations, $F\to\alpha^{-d}F$, as we have discussed in the introduction. For any metric corresponding to an extremum of $F$ one therefore finds that $T^{(F)}_{\hat\mu\hat\nu}$ vanishes. We may ask if an extremum of $F$ is also an extremum of $\Gamma$. In this case the field equations are the same as in the the absence of $F$. However, the possible solutions should be, in addition, extrema of $F$. This may impose conditions on the most general solution of the field equations in the absence of $F$. For the example of an extremum of $\Gamma$ with constant $\xi$ the extremum of $F$ should occur for a Ricci flat geometry. 

We recall, however, that there is no need for an extremum of $\Gamma$ to be an extremum of $F$. In the appendix we discuss an explicit example of a solution with $\hat R\hmn\neq 0$ for which $T^{(F)}\hmn$ does not vanish. Nevertheless, the extremum of the effective action occurs for $\Gamma_0=0~,~F_0=0$ and therefore belongs to the flat phase with a vanishing four-dimensional cosmological constant. The example admits an $SO(E+1)$ isometry, resulting in a non-abelian gauge symmetry of the effective four-dimensional theory. This demonstrates incidentally that solutions with non-abelian gauge symmetries are compatible with a dilatation symmetric effective action that leads to $\Lambda=0$ independently of the precise choice of parameters.

\section{Dilatation anomaly}
\label{dilatationanomaly}
If a dilatation symmetric ultraviolet fixed point exists and is reached for $\xi\to\infty$, the quantum  effective action may still contain terms that violate dilatation symmetry for finite values of $\xi$. From the point of view of the fixed point they arise from  small deviations from the exact fixed point corresponding to ``relevant'' or ``marginal'' directions in a renormalization group language. We may interprete such terms as a ``dilatation anomaly'', since the quantum theory exhibits an explicit breaking of the dilatation symmetry. For $\xi\to\infty$ the dilatation anomaly vanishes. In this section we discuss a particularly simple setting how a dilatation anomaly may be generated by quantum fluctuations.

For the understanding of the issue of a possible dilatation anomaly we first discuss a Weyl scaling of the higher dimensional effective action. This amounts to field rededications
\begin{equation}\label{4}
\hat g_{\hat\mu\hat\nu} =w^2\tilde g_{\hat\mu\hat\nu}~,~w=M_d\xi^{-\frac{2}{d-2}}
\end{equation}
and
\begin{equation}\label{9a}
\delta=\left(\zeta+\frac{4f_d}{(d-2)^2}\right)^{1/2}\ln(\xi/M_d^{(d-2)/2})
\end{equation}
Expressed in terms of $\tilde g\hmn$ and $\delta$ a scale invariant effective action
\begin{equation}\label{10a}
\Gamma=\frac{M^{d-2}_d}{2}\int\tilde g^{1/2}
\{-\tilde R+\partial^{\hat\mu}\delta~\partial_{\hat\mu}\delta\}.
\end{equation}
(which corresponds to eq. \eqref{AA1a} with $F=0$) becomes
\begin{equation}\label{1}
\Gamma=\int\hat g^{1/2}\left\{-\frac12\xi^2\hat R+\frac\zeta 2\partial^{\hat\mu}\xi\partial_{\hat\mu}\xi\right\}.
\end{equation}

After Weyl scaling, the effective action \eqref{10a} depends on $\delta$ only through derivative terms. This generalizes for all dilatation symmetric actions, as for example eq. \eqref{AA1a} with $F(\hat R_{\hat\mu\hat\nu\hat\rho\hat\sigma})\neq 0$. The action exhibits now a symmetry with respect to global shifts
\begin{equation}\label{ASA}
\delta\to\delta+c_\delta.
\end{equation}
Those correspond to the original dilatations which result in a global multiplicative rescaling of $\xi$. The field $\delta$ is the Goldstone boson of spontaneously broken dilatation symmetry. As it should be, it can only have derivative couplings.

If we would consider eq. \eqref{10a} as a classical action and compute quantum fluctuations, the global shift symmetry would be preserved and the effective action would necessarily be invariant under the shift \eqref{ASA}. This follows from the simple observation that a variable transformation in the functional integral introduces no Jacobian for a linear variable shift if the type \eqref{ASA}. The shift symmetry is therefore free of anomalies. The definition of a fundamental theory along these lines results in a dilatation symmetric quantum field theory without dilatation anomalies. Such a combination is always possible, establishing the possibility to have a dilatation invariant functional measure. 

For a fundamental theory without explicit mass scale it seems more natural, however, to employ the action \eqref{1} or \eqref{AA1a} for a starting point of a computation of quantum effects via a functional integral \cite{CWCC}. The functional measure is then formulated in terms of $\xi$ and $\hat g_{\hat\mu\hat\nu}$. Performing a variable transformation with a Weyl scaling \eqref{4} involves a non-trivial Jacobian. This is responsible for the dilatation anomaly.

The Jacobian $J$ can be computed explicitly. Since the variable $\xi$ remains unchanged, and the factor $w^2$ in eq. \eqref{4} depends on $\xi$ but not on the metric, one finds formally
\begin{equation}\label{K2}
J=\prod_{\hat x}[w^2\big(\xi(\hat x)\big)]^{\tilde f},
\end{equation}
where the product goes over all space-time points and $\tilde f$ counts the effective number of metric degrees of freedom. Writing the functional integral as
\begin{eqnarray}\label{K3}
Z=\int {\cal D}\xi{\cal D}\hat g_{\hat\mu\hat\nu} e^{-S}&=&
\int {\cal D}\xi{\cal D}\tilde g_{\hat\mu\hat\nu} e^{-(S+S^{(1)}_{an})}
\end{eqnarray}
the action after Weyl scaling acquires an anomalous piece
\begin{equation}\label{K4}
S^{(1)}_{an}=-\ln J=-\tilde f\sum_{\hat x}\ln w^2.
\end{equation}
(Here we use ``euclidean conventions'' for the action and functional integral, such that $\Gamma=S~+$ quantum corrections. Appropriate analytic continuation between Minkowski and euclidean signature is understood.) 

A second anomalous piece arises from the variable transformation from $\xi$ to $\delta$ by eq. \eqref{9a}
\begin{eqnarray}\label{K5}
Z&=&\int{\cal D}\delta{\cal D}\tilde g_{\hat\mu\hat\nu} e^{-(S+S_{an})},\\
S_{an}&=&S^{(1)}_{an}+S^{(2)}_{an},\nonumber\\
S^{(2)}_{an}&=&-\sum_{\hat x}\ln
\left(\frac{\kappa\xi}{M^{(d-2)/2}_d}\right)~,~
\kappa=\left(\zeta+\frac{4f_d}{(d-2)^2}\right)^{-1/2}.\nonumber
\end{eqnarray}
We can combine the anomalous pieces as
\begin{equation}\label{K6}
S_{an}=\frac f2\sum_{\hat x}\big[\ln(\xi^2/M^{d-2}_d)+\textup{const}\big],
\end{equation}
with $f>0$ since $S^{(1)}_{an}$ dominates over the $\xi$-dependent part.

We have to regularize the sum over all space-time points in eq. \eqref{K6}. This regularization should preserve diffeomorphism invariance which enforces $\sum_{\hat x}$ to be proportional to the invariant volume
\begin{equation}\label{K7}
\sum_{\hat x}=\mu^d\int_{\hat x}\hat g^{1/2}.
\end{equation}
The scale $\mu$ indicates that the Jacobian of the variable transformation violates the scaling symmetry. It is necessary for dimensional reasons since $\int_{\hat x}\hat g^{1/2}$ scales as a volume $\sim$ mass$^{-d}$, and $\sum_{\hat x}$ is dimensionless. The length scale $\mu^{-1}$ sets the units in which the continuous variable $\hat x$ is ``measured''. 

Employing eq. \eqref{K7} the anomalous piece in the action reads
\begin{equation}\label{K8}
S_{an}=\frac f2 \mu^d\int \hat g^{1/2}
\big[\ln(\xi^2/M^{d-2}_d)+\textup{const}\big].
\end{equation}
By the Weyl scaling \eqref{4} it can be expressed in terms of $\tilde g_{\hat\mu\hat\nu}$ and $\delta$ as 
\begin{eqnarray}\label{K9}
S_{an}=f\kappa \mu^d\int \tilde g^{1/2}\exp 
\left(-\frac{2d\kappa}{d-2}\delta\right)(\delta+c_\mu),
\end{eqnarray}
with $c_\mu$ an appropriate constant. As it should be, the scale $\mu$ is not observable. It can be rescaled arbitrarily by a shift in $\delta$ (together with a change in the constant $c_\mu$). 

For the final functional integral over $\tilde g_{\hat\mu\hat\nu}$ and $\delta$ in eq. \eqref{K3} we therefore have to add to the shift invariant action \eqref{10a} the anomalous action $S_{an}$ \eqref{K9}. Instead of a free massless Goldstone boson the field $\delta$ describes now a pseudo-Goldstone boson, with potential
\begin{equation}\label{K10}
V(\delta)=f\kappa\mu^d\exp \left(-\frac{2d\kappa}{d-2}\delta\right)(\delta+c_\mu).
\end{equation}
Its effective mass term is given by the second derivative with respect to $\delta$
\begin{equation}\label{K11}
\frac{m^2_\delta}{M^2_d}=\frac12 M^{-d}_d\frac{\partial^2V}{\partial\delta^2}\sim
\left(\frac{\mu}{M_d}\right)^d\exp\left(-\frac{2d\kappa}{d-2}\delta\right).
\end{equation}
We can choose the value of the scale of spontaneous dilatation symmetry breaking $M_d$ such that in the range of interest $\delta$ is close to zero. We find then for the mass of the pseudo Goldstone boson
\begin{equation}\label{K12}
m^2_\delta\sim \mu^d/M^{d-2}.
\end{equation}

For a small mass scale of the anomaly $\mu$ as compared to the spontaneous symmetry breaking scale $M_d$ the mass of the pseudo Goldstone boson can be strongly suppressed. (For $d=4$ one recovers the relation familiar for the axion and other pseudo Goldstone bosons, $m=\mu^2/M$.) In the limit $m_\delta\to 0$ the functional integral  over $\tilde g_{\hat\mu\hat\nu}$ and $\delta$ preserves the shift symmetry \eqref{ASA} and therefore produces no further dilatation anomaly. For small $m^2_\delta$ we can expand the propagators in loops in powers of $m^2_\delta$. This generates additional contributions to the dilatation anomaly. The lowest order in this expansion will contribute to $V$ a term $\sim m^2_\delta M^{d-2}_d$ and only ``renormalizes'' the constants appearing in eq. \eqref{K10}. Higher orders in $m^2_\delta$ are further suppressed. We will therefore take eq. \eqref{K8} or \eqref{K9} as our final form of the dilatation anomaly. 

In this context it may be of interest that the generic form \eqref{K6} can also be obtained from a simple one loop calculation using the action \eqref{1} as a starting point. The inverse propagator $P$, as given by the second functional derivative of the action, has the generic form 
\begin{equation}\label{K13}
P=\left(\begin{array}{lll}
\xi^2\tilde P_g&,&\xi\tilde P_{\xi g}\\
\xi\tilde P_{\xi g}&,&\tilde P_\xi
\end{array}\right)
\end{equation}
with $\tilde P_g,\tilde P_{\xi,g}$ and $\tilde P_\xi$ depending on the metric and involving derivative operators, but independent of $\xi$. The one loop expression takes the generic form 
\begin{eqnarray}\label{K14}
S^{(1\textup{loop)}}&=&-\ln \prod_{\hat x}(\det P)^{-1/2}\nonumber\\
&=&\frac f2\sum_{\hat x}
\big(\ln\xi^2+{\cal L}[\hat g_{\hat\mu\hat\nu}]\big).
\end{eqnarray}
Here $f$ can be associated with the effective number of metric degrees of freedom. The simple form of the $\xi$-dependence of the one loop expression obtains since powers of $\xi$ can easily be factorized out in $\det P$.

\section{Cosmological runaway and \newline dark energy}
\label{cosmologicalrunaway}

The dilatation anomaly adds to the effective four dimensional action a contribution to the potential $V$
\begin{equation}\label{K15}
V_{an}=\tilde \mu^d\varphi^{-D}_l
\big[\ln(\varphi^2_\xi\varphi^D_l/\tilde \mu^{d-2})+v_0\big],
\end{equation}
where $l^D=\varphi^{-D}_l$ arises from the volume of internal space and we have absorbed constants into a redefinition of the anomaly scale $\tilde\mu$. For $\tilde F=\tilde Q=0$, as required by asymptotic solutions in the flat phase, this is the only contribution to $V$. 

If we assume that a cosmological solution settles to a constant ratio $\varphi_\xi/\varphi_l$ the potential decays for large $\varphi_\xi$
\begin{equation}\label{K16}
V\sim\tilde\mu^d\varphi^{-D}_\xi.
\end{equation}
Thus $\varphi_\xi$ will not reach a stable value and we expect cosmological solutions where $\varphi_\xi$ moves to infinity. For large time the dimensionless ratio
\begin{equation}\label{K17}
\frac{V}{\varphi^4_\xi}\sim \left(\frac{\mu}{\varphi_\xi}\right)^d
\end{equation}
can become very small. In the asymptotic limit, the anomaly $\sim \mu$ can be neglected. In this limit we may consider the solution as static, since also the ``driving force'' for the motion of $\varphi_\xi$, i.e. the derivative $\partial V/\partial\varphi_\xi$, becomes very small. In the asymptotic limit we can then neglect the dilatation anomaly and look for quasistatic solutions for an effective action with dilatation symmetry. This is precisely the setting for the investigations of the present paper. 

For finite cosmological time a cosmological solution will only approach the dilatation symmetric asymptotic solution. If we associate $\varphi_\xi$ (or more generally $\chi$) with the present value of the (reduced) Planck mass, the present value of $V$ sets the scale of the dark energy \cite{CWQ}. For a demonstration of the leading behavior we neglect the logarithmic dependence,
\begin{equation}\label{267}
V=\left(\frac{\bar\mu}{\varphi_\xi}\right)^d\varphi^4_\xi.
\end{equation}
This corresponds to an anomaly which is a simple higher dimensional cosmological constant \cite{CWQ}. After a Weyl scaling to the four dimensional Einstein frame the effective potential reads $(\tilde F=\tilde Q=\tilde G=0)$ 
\begin{equation}\label{268}
V_4=\left(\frac{\bar\mu}{\varphi_\xi}\right)^dM^4.
\end{equation}
A kinetic term for $\varphi_\xi$ of the form \eqref{B10a} (with $\partial_\mu\omega=0$) becomes in the Einstein frame \cite{CWQ} 
\begin{equation}\label{269}
L_{kin,4}=\frac{M^2}{2}(Z_\xi+6)\partial^\mu\ln
\left(\frac{\varphi_\xi}{\bar\mu}\right)\partial_\mu\ln
\left(\frac{\varphi_\xi}{\bar\mu}\right).
\end{equation}

The cosmon field $\varphi$ with a canonic kinetic energy is proportional $M\ln(\varphi_\xi/M)$
\begin{equation}\label{270}
\varphi=\sqrt{Z_\xi+6}M\ln
\left(\frac{\varphi_\xi}{\bar\mu}\right).
\end{equation}
In consequence, the cosmon potential \eqref{268} has an exponential shape
\begin{eqnarray}\label{271}
V_4&=&M^4\exp\left(-\alpha\frac{\varphi}{M}\right),\nonumber\\
\alpha&=&\frac{d}{\sqrt{Z_\xi+6}}.
\end{eqnarray}
We note that $\alpha$ may take values substantially larger than one, in particular for negative values of $Z_\xi$ close to the ``conformal value'' $Z_\xi=-6$. Exponential potentials have been the first candidates for a dynamical dark energy or quintessence and lead to cosmological scaling solutions \cite{CWQ} where $V_4$ and the dark energy density decrease $\sim t^{-2}$. This could explain why dark energy is of the same order as dark matter. The fraction in homogeneous early dark  energy is given by \cite{CWQ} (for matter dominating radiation)
\begin{equation}\label{272}
\Omega_h=\frac{3}{\alpha^2}.
\end{equation}
For realistic cosmologies such a scaling solution has to be ended by some cosmological event, as for growing neutrino quintessence \cite{CWN}. 

We observe that an additional logarithmic dependence as in eq. \eqref{K15} multiplies the cosmon potential by a factor
\begin{eqnarray}\label{273}
A&=&\bar v+(d-2)\ln(\varphi_\xi/\bar\mu)\nonumber\\
&=&\bar v+\frac{d-2}{\sqrt{Z_\xi+6}}\frac{\varphi}{M}=\bar v+
\left(1-\frac 2d\right)\alpha\frac{\varphi}{M}.
\end{eqnarray}
The cosmological effects of such a factor may best be visualized by defining a $\varphi$-dependent $\alpha(\varphi)$
\begin{eqnarray}\label{274}
V_4&=&AM^4\exp \left(-\alpha\frac{\varphi}{M}\right)=M^4\exp
\left(-\alpha(\varphi)\frac{\varphi}{M}\right),\\
\alpha(\varphi)&=&\alpha-\frac{M}{\varphi}\ln A(\varphi)=
\alpha-\frac{M}{\varphi}\ln
\left\{\bar v+\left(1-\frac{2}{d}\right)\alpha\frac{\varphi}{M}\right\}.\nonumber
\end{eqnarray}
This leads to a decrease of $\alpha(\varphi)$ as $\varphi$ increases with cosmological time, and therefore to an increase of the dark energy fraction $\Omega_h$. Realistic present values of the dark energy density require a present value of the cosmon field given by $\alpha(\varphi)\varphi/M\approx 276$. This implies $\alpha(\varphi)\approx \alpha$ up to small corrections. 

We may define by $\alpha_0$ the present value of $\alpha$ and by $\varphi_0$ the present cosmon field. Using $\alpha_0\varphi_0/M=276$ we can give the quantitative value of the scale $\bar\mu$ which characterizes the dilatation anomaly
\begin{eqnarray}\label{275}
\left(\frac{\bar\mu}{\varphi_{\xi,0}}\right)^d&=&
\left(\frac{\bar\mu}{M}\right)^d=
\frac{V_0}{M^4}=10^{-120},\nonumber\\
\bar\mu&=&10^{-\frac{120}{d}}M\approx 10^{18-\frac{120}{d}}GeV.
\end{eqnarray}
For $d=10(18)$ this would amount to a high energy scale $\bar\mu=10^6(10^{11})$ GeV.

\section{Conclusions}
\label{conclusions}

In general, the quantum effective action $\Gamma[\hat g_{\hat\mu\hat\nu},\xi]$ for gravity coupled to a scalar field $\xi$ has a complicated form. We explore here the possibility of a simple scaling limit for large $\xi$ and small 
$\hat g_{\hat\mu\hat\nu}$. More precisely, we investigate the scaling with a multiplicative power of a dimensionless parameter $\kappa$ and define
\begin{equation}\label{Z1}
\Gamma_\kappa[\hat g_{\hat\mu\hat\nu},\xi]=\Gamma[\kappa^{-2}\hat g_{\hat\mu\hat\nu},\kappa^{\frac{d-2}{2}}\xi].
\end{equation}
Our hypothesis is a simple limiting behavior of $\Gamma_\kappa$ for $\kappa\to\infty$. In particular, we investigate a possible fixed point \cite{CWQ}, \cite{CWNL}
\begin{eqnarray}\label{Z2}
\lim_{\kappa\to\infty}\Gamma_\kappa=\int_{\hat x}\hat g^{1/2}
\left\{-\frac12\xi^2\hat R+\frac{\zeta}{2}
\partial^{\hat \mu}\xi\partial_{\hat \mu}\xi\right\}.
\end{eqnarray}
As appropriate for a fixed point, eq. \eqref{Z2} states that $\Gamma_\kappa$ becomes independent of $\kappa$ for $\kappa\to\infty$. In this limit the effective action is therefore invariant with respect to dilatations. 

If a fixed point \eqref{Z2} exists, it may be used for a definition of quantum gravity, with a non-perturbative renormalization similar to the asymptotic safety scenario \cite{25}, \cite{26}. Quantum fluctuations are expected to modify the simple asymptotic form of the effective action by adding for finite $\kappa$ additional terms to the r.h.s. of eq. \eqref{Z2}. Such terms typically violate the dilatation symmetry and will then be treated as ``dilatation anomalies''. An example would be a cosmological constant
\begin{equation}\label{Z3}
\Gamma^{(\bar\mu)}=\int_{\hat x}
\hat g^{1/2}\bar\mu^d.
\end{equation}
Its contribution to $\Gamma_\kappa$ indeed vanishes for $\kappa\to\infty$
\begin{equation}\label{Z4}
\Gamma^{(\bar\mu)}_\kappa=\kappa^{-d}\int_{\hat x}\hat g^{1/2}\bar\mu^d.
\end{equation}

One may imagine a non-linear functional flow equation for the $\kappa$-dependence of $\Gamma_\kappa$ (at fixed $\hat g_{\hat\mu\hat\nu},\xi$ and besides the trivial linear one following from the definition \eqref{Z1}),
\begin{equation}\label{Z5}
\kappa\partial_\kappa\Gamma_\kappa[\hat g_{\hat\mu\hat\nu},\xi]=
{\cal F}[\hat g_{\hat\mu\hat\nu},\xi],
\end{equation}
with a functional ${\cal F}$ typically involving $\Gamma_\kappa$ and its functional derivatives. One could then define quantum gravity by a solution of this flow equation with the ``boundary condition'' or ``initial value'' \eqref{Z2}. The fixed point would be an ``ultraviolet fixed point'' in the sense of $\kappa\to\infty$. Effective actions which are close to this fixed point for finite $\kappa$ may then be treated in the standard renormalization group formalism for relevant and marginal directions.

The fixed point may become important for late time cosmological solutions. This happens if $\xi(t)$ increases and $\hat g^{1/2}(t)$ decreases for increasing time, such that the field equations, which obtain from the functional derivatives of $\Gamma$ with respect to $\hat g_{\hat\mu\hat\nu}$ and $\xi$, have to be evaluated for values of the fields where the limit $\Gamma_{\kappa\to\infty}$ becomes relevant. In sect. \ref{cosmologicalrunaway} we have discussed an explicit example for this type of ``runaway cosmology''. The approach to the fixed point leads to an interesting candidate for dynamical dark energy, where the potential energy of an appropriate scalar field vanishes only asymptotically for $t\to\infty$. 

The fixed point \eqref{Z2} may exist for arbitrary dimension $d>2$, including $d=4$. In this paper we are interested in higher dimensional theories. Indeed, for $d>6$ no polynomial potential for $\xi$ is consistent with dilatation symmetry. This may add to the plausibility of a fixed point effective action without a potential $V(\xi)$. On the other hand, the vanishing of the higher dimensional cosmological constant is, in general, not sufficient to guarantee a vanishing effective four-dimensional cosmological constant after ``spontaneous compactification'' of the internal dimensions. The curvature of internal space or a non-trivial warping may generate such an effective cosmological constant. One of the important findings of the present paper is a statement about the most general ``compactified'' higher dimensional solutions of the field equations derived from the fixed point action \eqref{Z2}: all stable solutions with maximal four-dimensional symmetry and finite effective four-dimensional gravitational constant must have a vanishing four-dimensional cosmological constant. This asymptotic vanishing of the cosmological constant can be understood by a self-adjustment mechanism for infinitely many degrees of freedom, as discussed in sect. \ref{adjustment}.

Consider now stable cosmological runaway solutions which drive the fields into a region where the fixed point effective action \eqref{Z2} becomes relevant, and which allow for an effective four-dimensional gravity. Such solutions lead to a vanishing four-dimensional cosmological constant for $t\to\infty$, thus solving the ``cosmological constant problem''. Furthermore, since the Universe is very old (in units of the Planck time), but not infinitely old, a small amount of homogeneous energy density remains present in the effective four-dimensional Universe. This could give an explanation for the observed dark energy. Due to the huge age of the Universe (in Planck units) we expect that the present geometry of the Universe is very close to the asymptotic geometry, which is a quasi static solution of the field equations derived from eq. \eqref{Z2}.

It is easy to find candidates for such asymptotic solutions - the simplest being a direct product of four-dimensional flat space and an internal $d-4$-dimensional Ricci-flat space with finite volume, accompanied by a constant value of the scalar field $\xi$. If internal space admits isometries, the gauge couplings of the resulting four-dimensional gauge interactions become time-independent for large $t$. This solution has a vanishing four-dimensional cosmological constant $\Lambda$ and exists for all values of $\zeta$ in eq. \eqref{Z2}. On the other hand, we have shown that all possible stable quasistatic solutions of the field equations derived from the effective action \eqref{Z2}, which are consistent with four dimensional gravity, must have a vanishing four-dimensional constant $\Lambda=0$. Stable asymptotic solutions approaching de Sitter or anti-de Sitter space $(\Lambda\neq 0)$ are not possible. This singles out a vanishing cosmological constant for the asymptotic behavior for $t\to\infty$ and may be the basis for the solution of the ``cosmological constant problem''.

Furthermore, we have extended our discussion to a more general form of a fixed point effective action. It includes dilatation symmetric higher curvature invariants for the metric. In the absence of a potential for $\xi$ our conclusions remain essentially unchanged. Finally, we have presented a simple estimate of the dilatation anomaly and discussed the qualitative behavior of the corresponding runaway cosmology. It approaches a zero cosmological constant for infinite time, and has a dynamical dark energy component of similar size as dark matter.

Our main conclusion is simple: if the quantum effective action for large $\xi$ and small 
$\hat g^{1/2}$ shows a limiting fixed point behavior \eqref{Z2}, or a similar dilatation symmetric  behavior without a potential term for $\xi$, and if a stable cosmological runaway solution approaches this field region for large $t$, the cosmological constant problem can be solved without any fine tuning of parameters or initial conditions.

\LARGE
\section*{APPENDIX A: CURVED GEOMETRIES WITH VANISHING COSMOLOGICAL CONSTANT}
\renewcommand{\theequation}{A.\arabic{equation}}
\setcounter{equation}{0}

\normalsize
In this appendix we present an instructive example for a geometry with a non-vanishing Ricci tensor $\hat R_{\hat\mu\hat\nu}$, that nevertheless implies a vanishing cosmological constant. We restrict $F$ to be a polynomial of $\hat R,\hat H_{\hat\mu\hat\nu}$ and their covariant derivatives. We have already found that in this case all Ricci-flat geometries ${\cal R}^d$ are acceptable extrema,
\begin{equation}\label{G1}
\hat R=0~,~\hat H_{\hat\mu\hat\nu}=0~,~\xi=\xi_0.
\end{equation}
With $R^d={\cal M}^4\times {\cal R}^D$, and ${\cal R}^D$ a Ricci flat $D$-dimensional space with finite volume, the effective four-dimensional physics exhibits finite nonzero values for the gravitational constant, as well as for the gauge couplings in case of isometries. In this appendix we are interested in possible additional solutions with non-Ricci-flat geometries, where
\begin{equation}\label{G2}
\hat H_{\hat\mu\hat\nu}\neq 0~,~\hat R=0~,~\xi=\xi_0.
\end{equation}

Let us consider $d=2~mod~4$ and write
\begin{equation}\label{G3}
F=f_1(\hat R^2,H_2)\hat R+f_2(\hat R^2,H_2)H_3+F_3+F_4,
\end{equation}
with
\begin{equation}\label{G4}
H_2=H_{\hat\mu\hat\nu}\hat H^{\hat\nu\hat\mu}~,~H_3=\hat H_{\hat \mu} \ ^{\hat\nu}\hat H_{\hat\nu} \ ^{\hat\rho}\hat H_{\hat\rho} \ ^{\hat\mu}.
\end{equation}
The term $F_3$ contains contractions of the indices of $\hat H_{\hat\mu\hat\nu}$ where at least four powers of $\hat H$ are involved, like $H_4=\hat H_{\hat \mu} ^{\hat\nu}\hat H_{\hat\nu}^{\hat\rho}\hat H_{\hat\rho}^{\hat\sigma}
\hat H_{\hat\sigma}^{\hat\mu}$, or terms involving higher powers of $H_3$. Finally, $F_4$ contains covariant derivatives of $\hat R$ and $\hat H_{\hat\mu\hat\nu}$. In $d=2~mod~4$ a term involving only $H_2$ is forbidden. The invariant $H_2$ has dimension mass$^4$, and a polynomial involving only $H_2$ scales therefore $\sim$ mass$^{4n}~,~n\in {\mathbbm N}$. Dilatation symmetry requires that any polynomial scales $\sim$ mass$^d$. For $d=2~mod~4$ at least one term with dimension mass $2m,m$ odd, is needed, as $\hat R$ or $H_3$, or covariant derivatives must be involved. The dimensionless functions $f_1$ and $f_2$ contain $p$ powers of $H_2$ and $q$ powers of $\hat R^2$, with $p+q=(d-2)/4$ for $f_1$ and $p+q+1=(d-2)/4$ for $f_2$. The form \eqref{G3} for $F$ constitutes for $d=2~mod~4$ the most general polynomial dilatation symmetric effective action not involving $\hat C_{\hat\mu\hat\nu\hat\rho\hat\sigma}$. 

As an example of an extremum that leads to a vanishing four-dimensional cosmological constant we consider the direct product
\begin{equation}\label{G5}
{\cal M}^4\times {\cal R}^{D_1-4}\times S^E\times N^E,
\end{equation}
with ${\cal R}^{D_1-4}$ a Ricci-flat $D_1-4$ dimensional space $(D_1=d-2E)$ with finite volume, while $S^E$ is the $E$-dimensional sphere with radius $a$ and Ricci tensor obeying
\begin{equation}\label{G6}
\hat R_{\bar\gamma\bar\delta}=\frac{E-1}{a^2}\hat g_{\bar\gamma\bar\delta}.
\end{equation}
The $E$-dimensional space $N^E$ has negative curvature with Ricci tensor opposite to $S^E$,
\begin{equation}\label{G7}
\hat R_{\bar\epsilon\bar\eta}=-\frac{E-1}{a^2}\hat g_{\bar\epsilon\bar\eta}.
\end{equation}

For this ansatz, the curvature scalar vanishes, $\hat R=0$, while $\hat H_{\hat\mu\hat\nu}=\hat R_{\hat\mu\hat\nu}$ remains different from zero. A simple computation yields
\begin{equation}\label{G8}
H_2=\frac{2E(E-1)^2}{a^4}~,~H_3=0.
\end{equation}
Indeed, for $\hat H_{\hat\mu\hat\nu}$ we find nonzero values only for the $2E$-dimensional subspace $S^E\times N^E$, with indices $\gamma,\delta=1\dots 2E$ and $\hat g_{\gamma\delta}=diag(\hat g_{\bar\gamma\bar\delta},\hat g_{\bar\epsilon\bar\eta})$. Correspondingly, one obtains for the contraction $H_{\hat\mu}\ ^{\hat\rho}\hat H_{\hat\rho\hat\nu}$ the non-zero entries
\begin{equation}\label{G9}
\hat H_\gamma \ ^{\hat\rho}\hat H_{\hat\rho\delta}=\frac{(E-1)^2}{a^4}\hat g_{\gamma\delta}.
\end{equation}
This explains the vanishing of $H_{n>2}$. In consequence $F_3$ vanishes for the ansatz \eqref{G5}, as well as the pieces involving $f_1$ and $f_2$. Furthermore, all covariant derivatives of $\hat H_{\hat\mu\hat\nu}$ vanish, $D_{\hat\rho}\hat H_{\hat\mu\hat\nu}=0$, and $\partial_{\hat\rho}\hat R=\partial_{\hat\rho}H_2=0$. Thus $F_4$ also vanishes. If the ansatz \eqref{G5} solves the field equations, we are granted that $F(\hat R_{\hat\mu\hat\nu\hat\rho\hat\sigma})$ vanishes. The four dimensional cosmological constant must therefore be zero.

It remains to be shown that the ansatz \eqref{G5} solves the field equations. Since it corresponds to a regular geometry, solutions of the field equations are guaranteed to correspond to an extremum of the action. For $\xi=\xi_0~,~\hat R=0$, the scalar field equations are solved. For deriving the field equations for the metric we use the identities for variations
\begin{equation}\label{G10}
\delta(\hat g^{\hat\mu\hat\nu} \hat H_{\hat\nu\hat\mu})=\delta\hat H^{\hat\mu} \ _{\hat\mu}=\delta\hat H^\gamma\ _\gamma+\delta\hat H^\tau\ _\tau=0
\end{equation}
and
\begin{equation}\label{G11}
\delta H_3=3\hat H^{\hat\mu} \ _{\hat\rho}\hat H^{\hat\rho} \ _{\hat\nu}\delta\hat H^{\hat\nu} \ _{\hat\mu}
=\frac{3(E-1)^2}{a^4}\delta\hat H^\gamma \ _\gamma.
\end{equation}
Here the index $\tau$ stands for the coordinates of the Ricci-flat $d-2E$ dimensional subspace, and we note that $\delta\hat H^\tau\ _\tau$ does not vanish in general. 
Furthermore, the non-vanishing variations are (for $\hat R=0$)
\begin{eqnarray}\label{G12}
\frac{\delta\hat H^\gamma \ _\gamma}{\delta\hat g^{\sigma\tau}}=-
\frac{\delta\hat H^{\sigma'} \ _{\sigma'}}{\delta\hat g^{\sigma\tau}}=b_1\hat H_{\sigma\tau},\nonumber\\
\frac{\delta\hat H^{\gamma'} \ _{\gamma'}}{\delta\hat g^{\gamma\delta}}=-
\frac{\delta\hat H^\sigma \ _{\sigma}}{\delta\hat g^{\gamma\delta}}=b_2\hat H_{\gamma\delta}.
\end{eqnarray}
(For simplicity, we omit here the contributions from terms in $F_3$ which have the form 
$H_2^{\frac{d-2}{4}-k}H_{2k+1},~k\geq 2$, and therefore vanish for the ansatz \eqref{G5}.) This yields for the $(\sigma,~\tau)$-component of the gravitational field equations
\begin{equation}\label{G13}
\big[\xi^2_0-\left(2\bar f_1+\frac{3b_1\bar f_2}{2E}\right)H_2^{\frac{d-2}{4}}\big]\hat H_{\sigma\tau}=0, 
\end{equation}
where we have defined
\begin{equation}\label{G14}
f_1(\hat R=0,H_2)=\bar f_1 H^{\frac{d-2}{4}}_2~,~f_2(\hat R=0,H_2)=\bar f_2
H_2^{\frac{d-6}{4}}.
\end{equation}
The field equation \eqref{G13} is obeyed for the Ricci-flat subspace $M^4\times R^{D_1-4}$ since $\hat H_{\sigma\tau}=0$. 

Similarly, the field equations for the $(\gamma,\delta)$-components are
\begin{equation}\label{G15}
\left[\xi^2_0-\left(2\bar f_1+\frac{3b_2\bar f_2}{2E}\right)
H_2^{\frac{d-2}{4}}\right]\hat H_{\gamma\delta}=0.
\end{equation}
For a non-zero $\hat H_{\gamma\delta}$ the bracket has to vanish. This fixes $H_2$ and therefore the characteristic scale $a$ for the $2E$-dimensional subspace as a function of $\xi_0$, according to
\begin{equation}\label{G16}
H_2^{\frac{d-2}{4}}=\frac{2E\xi^2_0}{4E\bar f_1+3b_2\bar f_2}.
\end{equation}
Provided that $4E\bar f_1>-3b_2 \bar f_2$ the ansatz \eqref{G5} indeed solves the field equations. No tuning of parameters is necessary, and the cosmological constant vanishes for a wide range of parameters $\bar f_1,\bar f_2$. 

In order to have an acceptable four dimensional gravity we also need a finite non-zero value of the effective gravitational constant. With $\Omega_D$ the finite volume of the $D=d-4$ dimensional internal space and an expansion of $H_3$ in powers of the four-dimensional curvature tensor $R^{(4)}$, $H_3=H^{(0)}_3+b_3H^{(0)}_2R^{(4)}$, one finds
\begin{eqnarray}\label{G17}
\chi^2&=&\Omega_D\big[\xi^2_0-2(\bar f_1+b_3\bar f_2)(H^{(0)}_2)^{\frac{d-2}{4}}\big]\nonumber\\
&=&-\Omega_D\bar f_2(H^{(0)}_2)^{\frac{d-2}{4}}
\left(2b_3+\frac{3b_2}{2E}\right).
\end{eqnarray}
Positivity of $\chi^2$ restricts the sign of $\bar f_2$ according to the sign of the combination $2b_3+3b_2/(2E)$. Finiteness of $\chi^2$ requires finite $\Omega_D\xi^2_0$. The gauge couplings of the $SO(E+1)$ gauge symmetry corresponding to the isometries of $S^E$ are also finite and non-zero if $\Omega_D$ is finite.


\end{document}